\documentclass[
  aps,
  prd,
  11pt,
  superscriptaddress,
  notitlepage,
  onecolumn,
  nofootinbib
]{revtex4-1}

\AtBeginDocument{
  \linespread{1.05}\selectfont
}

\usepackage[
  a4paper,
  left=2.3cm,
  right=2.3cm,
  top=2.5cm,
  bottom=2.5cm
]{geometry}

\usepackage{microtype} 
\usepackage{graphicx}
\usepackage{xcolor}
\usepackage{orcidlink}
\usepackage{enumerate}
\usepackage{enumitem}
\usepackage{ulem}
\usepackage{amssymb} 
\usepackage[svgnames]{xcolor} 
\usepackage{tikz}
\usepackage{mathtools}
\usepackage[dvipsnames]{xcolor}
\usepackage{ytableau}
\usepackage{soul}
\usetikzlibrary{arrows.meta,positioning,fit,calc}

\definecolor{linkcolor}{rgb}{0.0,0.3,0.5}
\definecolor{linkcolor}{rgb}{0.0,0.3,0.5}
\definecolor{mypurple}{RGB}{143, 116, 210}

\newcommand{\iu}{\mathrm{i}\mkern1mu}

\graphicspath{{Figures/}}

\newcommand{\hias}{School of Fundamental Physics and Mathematical Sciences, Hangzhou Institute for Advanced Study, University of Chinese Academy of Sciences, Hangzhou 310024, China}

\newcommand{\bnu}{School of Physics and Astronomy, Beijing Normal University, Beijing 100875, China}

\begin{document}

\title{Taiji Resolving Power for the Transverse Scalar Mode of Gravitational Waves}

\author{Bo-Xuan Ge
\orcidlink{0000-0003-0738-3473}}
\email{bo-xuan.ge@ucas.ac.cn}
\affiliation{\hias}
\author{Zhoujian Cao\footnote{corresponding author}
\orcidlink{0000-0002-1932-7295}}
\email{zjcao@bnu.edu.cn}
\affiliation{\hias}
\affiliation{\bnu}

\begin{abstract}
We investigate how small a co-propagating transverse-scalar
component can be resolved by Taiji in an already identified bright
tensor chirp.  Using a source-tracked tensor-null response, we
formulate the problem in terms of the minimum resolvable scalar strain
fraction and evaluate it over the sky.  For a one-year benchmark chirp with tensor signal-to-noise ratio
\(\rho_T=1000\), we find that Taiji can rule out transverse-scalar
strain fractions \(\epsilon_b\gtrsim0.532\%\) at the all-sky median
level.  The threshold scales
as \(\epsilon_{b,\min}\propto\rho_T^{-1}\), so brighter tensor events
can probe correspondingly smaller scalar fractions.  We further find
that this sub-percent resolving power remains predictable under small
source-parameter mismatches through the associated tensor-leakage
structure.
\end{abstract}


\maketitle

\section{Introduction}
\label{sec:introduction}

Gravitational-wave observations provide a direct test of the
propagating degrees of freedom of gravity.  In general relativity
(GR), a gravitational wave in vacuum carries only the two
transverse-traceless tensor polarizations, conventionally denoted by
\(+\) and \(\times\).  A signal containing an additional polarization
would therefore indicate gravitational dynamics beyond the tensor
sector of GR.  More general theories can admit scalar or vector
polarizations, and their interferometric signatures have long been
studied as a way of testing the polarization content of gravitational
waves \cite{Nishizawa:2009bf}.

Among these possibilities, a transverse-scalar polarization produces
an isotropic expansion and contraction in the plane orthogonal to the
propagation direction and is therefore commonly referred to as the
breathing mode.  The observation of such a component would not by
itself identify a unique theory of gravity, but it would indicate that
the two tensor polarizations of GR are insufficient to describe the
measured signal.  This motivates asking not only whether an
interferometer responds to a breathing polarization, but how small
such a component could be distinguished from an otherwise
tensor-dominated gravitational-wave signal.

Space-based interferometers are particularly suited to polarization
tests of long-lived signals.  Their long baselines, triangular
geometry, and orbital motion produce a detector response that changes
continuously with time and source direction.  The Taiji mission
provides access to this regime with a heliocentric three-spacecraft
constellation \cite{Hu:2017mde}.  For a chirping signal whose
time-frequency evolution can be followed over the observation, the
changing detector geometry can be combined with the source track to
construct polarization-sensitive combinations that evolve together
with the signal.

Compact-binary systems provide a natural class of such chirping
sources and have been studied extensively in a variety of settings,
including binary black holes, binary neutron stars, and binary boson
stars
\cite{
Pretorius:2005gq,
Campanelli:2005dd,
LIGOScientific:2016aoc,
Shibata:1999wm,
Baiotti:2008ra,
LIGOScientific:2017vwq,
Croft:2022bxq,
Evstafyeva:2022bpr,
Ge:2024fum,
Ge:2024itl,
Ge:2025btw,
Ge:2026wzh,
Ge:2026kkq,
Ning:2026qxs,
Marks:2026xvo,
Damour:2025oys,
Brito:2025rld,
Evstafyeva:2024qvp,
CaoBinary:CaoLiu2011HeadOn,
CaoBinary:HanCao2011IMRI,
CaoBinary:CaoHan2017EccentricEOBNR,
CaoBinary:LiuEtAl2024PrecessingEccentric,
CaoBinary:ZhaoEtAl2024EMRI,
CaoBinary:YueCao2024IMRIWaveforms}.
For an identified bright compact-binary signal, the dominant tensor
component can provide the source position and time-frequency track
needed to search for a weaker additional polarization in the same
data.

A transverse-scalar mode also arises in gravitational quantum field
theory (GQFT), which provides one concrete theoretical motivation for
the polarization considered here.  GQFT has been developed from a
spin-gauge formulation of gravity toward more general constructions
\cite{Wu:2015wwa,Wu:2024mul,Wu:2025abi}.  Its linearized
gravitational dynamics contains, in addition to the two tensor modes,
one scalar and two vector propagating degrees of freedom
\cite{Gao:2025aye}.  For the interferometric response relevant here,
the additional scalar degree of freedom appears as a transverse
breathing polarization.  The amplitude of this mode in a particular
GQFT source depends on the underlying source dynamics
\cite{Gao:2025aye}; in the present work, however, we do not assume a
specific source model for that amplitude.

The separation of an additional polarization from a much stronger
tensor signal is itself a detector problem.  Reference
\cite{Xu:2026kev} studied this problem for Taiji and LISA using
null-response channels (NRCs) constructed from the three Sagnac
combinations of a triangular interferometer.  In the tensor
null-response channel (t-NRC), the combination coefficients are chosen
so that the \(+\) and \(\times\) responses cancel while a breathing
response generally remains.  The construction depends on the detector
response rather than on a particular source amplitude, and the
time-dependent constellation can substantially modify the
breathing-mode response \cite{Xu:2026kev}.

The response of a null channel alone does not directly answer the
observational question relevant to an identified event.  Suppose that
a bright tensor chirp has already been detected, with its sky position
and time-frequency evolution sufficiently well determined.  The
quantity of interest is then the smallest additional breathing
component that can be resolved relative to that observed tensor
signal.  We characterize it by the strain-amplitude fraction
\begin{equation}
    \epsilon_b
    =
    \frac{H_b(t)}{H_T(t)},
\end{equation}
where \(H_T(t)\) is the polarization-averaged tensor
strain-amplitude envelope and \(H_b(t)\) is the corresponding
co-propagating breathing component.  The quantity
\(\epsilon_{b,\min}\) is therefore a detector-level resolving
threshold rather than a prediction for the scalar amplitude of a
particular theory.

In this work we extend the dynamic t-NRC construction to an evolving
source track and combine it with a consistent unequal-arm \(A/E\)
tensor normalization.  The tensor-null coefficients are recomputed
along the chirp as both the detector geometry and the
gravitational-wave frequency evolve.  The resulting scalar reach can
be expressed directly in terms of the tensor-network signal-to-noise
ratio \(\rho_T\), separating the brightness of the identified event
from the relative resolving power of the detector.  We evaluate the
corresponding threshold over an equal-solid-angle sky grid and examine
its dependence on the tensor SNR.

For the one-year benchmark chirp considered below, adopting
\(\rho_b^\star=5\) as an operational scalar-channel threshold, a
tensor event with \(\rho_T=1000\) gives an equal-solid-angle all-sky
median threshold
\begin{equation}
    \epsilon_{b,\min}^{50\%}
    \simeq
    0.532\%,
\end{equation}
with the most favorable directions reaching approximately \(0.31\%\).
For a fixed waveform track, sky position, observation time, and
detector configuration, the threshold scales as
\(\epsilon_{b,\min}\propto\rho_T^{-1}\).  The result therefore
quantifies the transverse-scalar fraction that can be tested in a
known bright event without assigning that fraction to a particular
source theory.

The ideal resolving power also depends on how accurately the tensor
signal is tracked.  Errors in the source position or chirp parameters
spoil the tensor cancellation and introduce residual tensor power into
the nominally null channel.  We therefore supplement the ideal
forecast with one-dimensional mismatch scans in the sky coordinates
and chirp parameters, followed by a local joint leakage metric that
captures parameter correlations and degeneracies.  The analysis thus
addresses both the minimum scalar fraction that can be resolved and
the source-tracking accuracy required to keep the residual tensor
leakage below the same scale.

Section~\ref{sec:null_scalar_fraction} introduces the dynamic
tensor-null response, the unequal-arm \(A/E\) tensor normalization,
and the scalar-fraction statistic.
Section~\ref{sec:scalar_reach} presents the benchmark chirp,
tensor-SNR scaling, and all-sky resolving power.
Section~\ref{sec:mismatch} studies tensor leakage caused by
source-parameter mismatch and its joint local structure.
The implications and limitations of the detector-level forecast are
discussed in Sec.~\ref{sec:discussion}.


\section{Dynamic null response and scalar-fraction statistic}
\label{sec:null_scalar_fraction}

We consider the two tensor polarizations of general relativity together
with a co-propagating transverse-scalar breathing mode.  Following the
polarization convention of Ref.~\cite{Xu:2026kev}, we introduce a
right-handed orthonormal basis
\((\hat{\mathbf k},\hat{\mathbf u},\hat{\mathbf v})\), where
\(\hat{\mathbf k}\) is the GW propagation direction, and define
\begin{align}
    \mathbf e_{+}
    &=
    \hat{\mathbf u}\otimes\hat{\mathbf u}
    -
    \hat{\mathbf v}\otimes\hat{\mathbf v},
    \\
    \mathbf e_{\times}
    &=
    \hat{\mathbf u}\otimes\hat{\mathbf v}
    +
    \hat{\mathbf v}\otimes\hat{\mathbf u},
    \\
    \mathbf e_b
    &=
    \hat{\mathbf u}\otimes\hat{\mathbf u}
    +
    \hat{\mathbf v}\otimes\hat{\mathbf v}.
\end{align}
Only these three polarizations are considered below.

\subsection{Dynamic tensor-null response}
\label{subsec:dynamic_null}

The detector response is built from the six directed one-way laser
links of the Taiji constellation.  For a unit-amplitude monochromatic
wave of polarization \(p\), the frequency-domain response of the link
from spacecraft \(i\) to spacecraft \(j\) can be written as
\cite{Xu:2026kev}
\begin{align}
    \mathcal Y_{ij}^{(p)}(f,t)
    &=
    \frac{
        \left(
            \hat{\mathbf r}_{ij}
            \otimes
            \hat{\mathbf r}_{ij}
        \right):\mathbf e_p
    }{
        2\left(
            1-\hat{\mathbf k}\cdot\hat{\mathbf r}_{ij}
        \right)
    }
    \left[
        \exp\left(
            -\frac{2\pi \iu f}{c}
            \hat{\mathbf k}\cdot\mathbf r_j
        \right)
        -
        \exp\left(
            -\frac{2\pi \iu f}{c}
            \left[
                L_{ij}
                +
                \hat{\mathbf k}\cdot\mathbf r_i
            \right]
        \right)
    \right],
    \label{eq:one_way_response}
\end{align}
where
\begin{equation}
    \hat{\mathbf r}_{ij}
    =
    \frac{\mathbf r_j-\mathbf r_i}{L_{ij}},
    \qquad
    L_{ij}
    =
    \left|\mathbf r_j-\mathbf r_i\right|.
\end{equation}
The spacecraft positions, arm directions, and arm lengths are
evaluated along the time-dependent orbit.  At each detector epoch, the
Sagnac variables \(\alpha\), \(\beta\), and \(\gamma\) are constructed
using the instantaneous first-generation TDI delay factors.  The
constellation geometry is therefore treated as frozen over the
light-travel time, while its orbital evolution and arm-length variation
are retained on the observation timescale.  Following
Ref.~\cite{Xu:2026kev}, we use the first-generation TDI response
functions.  Laser-frequency noise is assumed to be suppressed by the
TDI construction, while optical-metrology and acceleration noises are
retained explicitly.  For a flexing constellation, exact laser-noise
cancellation requires second-generation TDI; the corresponding
observables, however, have the same gravitational-wave sensitivity as
their first-generation counterparts \cite{Xu:2026kev}.  The results
below are therefore secondary-noise-limited resolving-power forecasts.

For each polarization, the Sagnac unit-strain responses are collected
into
\begin{equation}
    \mathbf R_p^{\rm Sag}(f,t)
    =
    \begin{pmatrix}
        \alpha_p(f,t)\\
        \beta_p(f,t)\\
        \gamma_p(f,t)
    \end{pmatrix},
    \qquad
    p=+,\times,b.
    \label{eq:sagnac_response_vector}
\end{equation}
The components are generally complex, with their moduli and phases
encoding the amplitude response and finite-light-travel-time phase
shifts, respectively.

Following the t-NRC construction of Ref.~\cite{Xu:2026kev}, we define
the tensor-null coefficients in the three-dimensional Sagnac-channel
space by
\begin{equation}
    a_t^{I}(f,t)
    =
    \epsilon^{IJK}
    R_{+,J}^{\rm Sag}(f,t)
    R_{\times,K}^{\rm Sag}(f,t),
    \qquad
    I,J,K\in\{\alpha,\beta,\gamma\},
    \label{eq:t_null_vector}
\end{equation}
up to an arbitrary nonzero complex normalization.  By antisymmetry of
\(\epsilon^{IJK}\), these coefficients satisfy
\begin{equation}
    a_t^{I}R_{+,I}^{\rm Sag}=0,
    \qquad
    a_t^{I}R_{\times,I}^{\rm Sag}=0,
    \label{eq:t_null_conditions}
\end{equation}
where repeated Sagnac-channel indices are summed.  Since the response
components are generally complex, Eqs.~\eqref{eq:t_null_vector} and
\eqref{eq:t_null_conditions} define a complex bilinear cancellation;
no complex conjugation is involved in the tensor-null condition.

For complex GW amplitudes \(\tilde h_p\), the signal contribution in
the Sagnac basis is
\begin{equation}
    \tilde s_I^{\rm Sag}
    =
    \tilde h_+ R_{+,I}^{\rm Sag}
    +
    \tilde h_\times R_{\times,I}^{\rm Sag}
    +
    \tilde h_b R_{b,I}^{\rm Sag}.
    \label{eq:sagnac_signal}
\end{equation}
The corresponding tensor-null channel is therefore
\begin{equation}
    \tilde\eta_t
    =
    a_t^I \tilde s_I^{\rm Sag}
    =
    \tilde h_b\,
    a_t^I R_{b,I}^{\rm Sag},
    \label{eq:t_null_signal}
\end{equation}
for an exactly specified source.  The two tensor contributions vanish,
whereas a breathing component remains whenever
\(a_t^I R_{b,I}^{\rm Sag}\neq0\).

Let \(\mathbf S_N^{\rm Sag}(f,t)\) denote the noise
cross-spectral-density matrix of the three Sagnac channels, with
\begin{equation}
    \left(S_N^{\rm Sag}\right)_{IJ}
    =
    \left\langle
        \tilde n_I^{\rm Sag}
        \tilde n_J^{{\rm Sag}*}
    \right\rangle .
    \label{eq:sagnac_noise_covariance}
\end{equation}
Since the tensor-null noise is combined with the same coefficients,
\(\tilde n_t=a_t^I\tilde n_I^{\rm Sag}\), its noise power is
\begin{equation}
    S_t(f,t)
    =
    a_t^I
    \left(S_N^{\rm Sag}\right)_{IJ}
    a_t^{J*}.
    \label{eq:t_null_noise}
\end{equation}

We then define the instantaneous noise-weighted breathing response
\begin{equation}
    K_b(f,t)
    =
    \frac{
        \left|
            a_t^I
            R_{b,I}^{\rm Sag}
        \right|^2
    }{
        a_t^I
        \left(S_N^{\rm Sag}\right)_{IJ}
        a_t^{J*}
    }.
    \label{eq:breathing_kernel}
\end{equation}
The numerator is the squared null-channel response to unit breathing
strain, while the denominator is the noise power of the same channel.
The ratio is invariant under
\(\mathbf a_t\rightarrow z\mathbf a_t\) for any nonzero
\(z\in\mathbb C\).

The t-NRC becomes poorly conditioned when the two tensor-response
vectors are nearly linearly dependent.  We monitor this with
\begin{equation}
    q_t(f,t)
    =
    \left[
        \frac{
            \left(
                \epsilon^{IJK}
                R_{+,J}^{\rm Sag}
                R_{\times,K}^{\rm Sag}
            \right)^*
            \left(
                \epsilon^{ILM}
                R_{+,L}^{\rm Sag}
                R_{\times,M}^{\rm Sag}
            \right)
        }{
            \left(
                R_{+,I}^{\rm Sag *}
                R_{+,I}^{\rm Sag}
            \right)
            \left(
                R_{\times,J}^{\rm Sag *}
                R_{\times,J}^{\rm Sag}
            \right)
        }
    \right]^{1/2}
    =
    \frac{
        \left\|
            \mathbf R_+^{\rm Sag}
            \times
            \mathbf R_\times^{\rm Sag}
        \right\|
    }{
        \left\|\mathbf R_+^{\rm Sag}\right\|
        \left\|\mathbf R_\times^{\rm Sag}\right\|
    }.
    \label{eq:tnrc_condition}
\end{equation}
where the norm is the standard Hermitian vector norm.  The primary
forecast retains epochs satisfying
\begin{equation}
    q_t\geq0.05.
    \label{eq:tnrc_condition_cut}
\end{equation}

For a chirping source, both the detector geometry and the GW frequency
vary during the observation.  The null coefficient is therefore
recomputed along the source track,
\begin{equation}
    \mathbf a_t(t)
    =
    \mathbf a_t[f(t),t],
    \label{eq:source_tracked_null}
\end{equation}
rather than being fixed at a reference epoch or frequency.

\subsection{Flexing-arm \(A/E\) tensor normalization}
\label{subsec:ae_tensor}

The tensor amplitude is normalized independently with the standard
Michelson-type TDI science channels.  Although the nominal
configuration has equal arm lengths, the heliocentric constellation
flexes during the orbit; the instantaneous lengths
\(L_{12}(t)\), \(L_{23}(t)\), and \(L_{31}(t)\) are therefore not
artificially constrained to be identical in the response calculation.

Starting from the same six one-way links, we construct the three
Michelson-type TDI variables \(X,Y,Z\).  They may be regarded as
synthetic Michelson interferometers with spacecraft \(1,2,3\),
respectively, as their reference vertices.  The three variables are
then transformed to the orthonormal combinations
\begin{align}
    A
    &=
    \frac{Z-X}{\sqrt{2}},
    \\
    E
    &=
    \frac{X-2Y+Z}{\sqrt{6}},
    \\
    T
    &=
    \frac{X+Y+Z}{\sqrt{3}}.
    \label{eq:aet_definition}
\end{align}
The \(A/E\) pair is used below as the tensor-sensitive science network;
the \(T\) combination is not included in the tensor normalization.

For polarization \(p\), its unit-strain response in the \(A/E\) basis
is written as
\begin{equation}
    \mathbf R_p^{AE}(f,t)
    =
    \begin{pmatrix}
        R_{A,p}(f,t)\\
        R_{E,p}(f,t)
    \end{pmatrix}.
    \label{eq:ae_response_vector}
\end{equation}
The corresponding noise is described by the full Hermitian covariance
matrix
\begin{equation}
    \mathbf C_{AE}(f,t)
    =
    \begin{pmatrix}
        S_{AA} & S_{AE}\\
        S_{EA} & S_{EE}
    \end{pmatrix},
    \qquad
    S_{EA}=S_{AE}^*.
    \label{eq:ae_covariance}
\end{equation}
We do not impose the equal-arm approximation
\(S_{AE}=0\).  The one-way optical-metrology and acceleration noises
are propagated through the same instantaneous arm-delay combinations
used for the GW response before transformation to the \(A/E\) basis.

The polarization-averaged tensor information rate is
\begin{equation}
    K_T(f,t)
    =
    \frac{1}{2}
    \sum_{p=+,\times}
    \left(R_{p,M}^{AE}\right)^*
    \left(C_{AE}^{-1}\right)^{MN}
    R_{p,N}^{AE}
    =
    \frac{1}{2}
    \sum_{p=+,\times}
    \left(
        \mathbf R_p^{AE}
    \right)^\dagger
    \mathbf C_{AE}^{-1}
    \mathbf R_p^{AE},
    \qquad
    M,N\in\{A,E\}.
    \label{eq:tensor_kernel}
\end{equation}
The factor \(1/2\) implements the tensor-polarization average.  This
definition provides the polarization-averaged tensor normalization
used throughout the scalar-fraction forecast.

\subsection{Scalar-fraction statistic}
\label{subsec:scalar_fraction}

We now separate the overall event brightness from the relative
amplitude evolution along the chirp.  Let
\begin{equation}
    H_T(t)
    =
    H_{T,0}\,\mathcal A(t),
    \qquad
    \mathcal A(0)=1,
    \label{eq:tensor_amplitude_envelope}
\end{equation}
denote the slowly varying tensor strain-amplitude scale associated
with the polarization-averaged \(A/E\) normalization of
Eq.~\eqref{eq:tensor_kernel}.  It is not the strain amplitude for a
specified source inclination or tensor-polarization state.
\(H_T(t)\) is an amplitude envelope rather than the full oscillatory
waveform; the rapidly varying GW phase is contained in the complex
signal amplitudes entering the detector response.

For the leading-order quasi-circular inspiral adopted here, the
tensor strain amplitude obeys the Newtonian scaling
\(H_T\propto f^{2/3}\).  We therefore take
\begin{equation}
    \mathcal A(t)
    =
    \left[
        \frac{f(t)}{f_0}
    \right]^{2/3},
    \label{eq:chirp_envelope}
\end{equation}
where \(f(t)\) is the GW frequency and \(f_0=f(0)\).

To formulate a detector-level scalar-fraction forecast, we introduce
a phenomenological co-propagating breathing component whose
strain-amplitude envelope is
\begin{equation}
    H_b(t)
    =
    \epsilon_b H_T(t).
    \label{eq:breathing_fraction_ansatz}
\end{equation}
Equivalently,
\begin{equation}
    \epsilon_b
    \equiv
    \frac{H_b(t)}{H_T(t)}.
    \label{eq:scalar_fraction_definition}
\end{equation}
The benchmark assumes that \(\epsilon_b\) is constant along the
observed track.  It is a strain-amplitude fraction, not an energy
fraction, an SNR fraction, or a fundamental coupling constant.

The breathing information accumulated by the source-tracked null
channel is
\begin{equation}
    I_b
    =
    \int_0^{T_{\rm obs}}
    dt\,
    \mathcal A^2(t)\,
    K_b[f(t),t],
    \label{eq:Ib}
\end{equation}
where epochs failing the conditioning requirement
Eq.~\eqref{eq:tnrc_condition_cut} are excluded.  The corresponding
tensor information accumulated in the \(A/E\) network is
\begin{equation}
    I_T
    =
    \int_0^{T_{\rm obs}}
    dt\,
    \mathcal A^2(t)\,
    K_T[f(t),t].
    \label{eq:IT}
\end{equation}

With the common SNR normalization used for the two information
integrals, the polarization-averaged tensor-network SNR is
\begin{equation}
    \rho_T^2
    =
    H_{T,0}^2 I_T,
    \label{eq:rho_tensor}
\end{equation}
while the breathing-mode SNR in the tensor-null channel is
\begin{equation}
    \rho_b^2
    =
    \epsilon_b^2
    H_{T,0}^2 I_b.
    \label{eq:rho_breathing}
\end{equation}
It follows that
\begin{equation}
    \frac{\rho_b}{\rho_T}
    =
    \epsilon_b
    \sqrt{\frac{I_b}{I_T}},
    \label{eq:snr_ratio}
\end{equation}
so that the strain fraction and SNR ratio are in general different.

Eliminating the unknown overall tensor amplitude \(H_{T,0}\), we
obtain
\begin{equation}
    \rho_b
    =
    \epsilon_b\,\rho_T
    \sqrt{\frac{I_b}{I_T}}.
    \label{eq:rho_b_rho_t}
\end{equation}
For an adopted scalar-channel SNR threshold
\(\rho_b^\star\), the minimum resolvable breathing fraction is
therefore
\begin{equation}
    \boxed{
    \epsilon_{b,\min}
    =
    \frac{\rho_b^\star}{\rho_T}
    \sqrt{\frac{I_T}{I_b}}
    }.
    \label{eq:epsilon_min}
\end{equation}

We adopt
\begin{equation}
    \rho_b^\star=5
\end{equation}
as the operational threshold for the forecasts below.  This value is
a reference choice rather than a universal detection threshold.
For fixed source track, sky position, observation time, and detector
response,
\begin{equation}
    \epsilon_{b,\min}
    \propto
    \frac{\rho_b^\star}{\rho_T}.
    \label{eq:epsilon_scaling}
\end{equation}

Equation~\eqref{eq:epsilon_min} separates the brightness of an
identified tensor event, represented by \(\rho_T\), from the
detector's relative resolving power for an additional breathing
component, encoded by \(I_T/I_b\).  The ideal construction assumes that
the source track entering \(\mathbf a_t[f(t),t]\) is known exactly.
Parameter errors spoil the tensor-null conditions and produce
residual tensor leakage; their effect is considered separately in
Sec.~\ref{sec:mismatch}.

\section{Scalar-fraction resolving power for a bright chirp}
\label{sec:scalar_reach}

We now apply the scalar-fraction statistic of
Sec.~\ref{sec:null_scalar_fraction} to a one-year chirping signal.
The calculation is conditional on an already identified tensor event:
its sky position and time-frequency track are assumed to be available
for constructing the source-tracked tensor-null channel, while the
amplitude of the co-propagating transverse-scalar component is left
free.

\subsection{Benchmark chirp}
\label{subsec:benchmark_chirp}

We use a leading-order quasi-circular inspiral track,
\begin{equation}
    f(t)
    =
    \left[
        f_0^{-8/3}
        -
        \frac{256}{5}
        \pi^{8/3}
        \left(
            \frac{G\mathcal M_c}{c^3}
        \right)^{5/3}
        t
    \right]^{-3/8},
    \label{eq:benchmark_chirp_frequency}
\end{equation}
together with the amplitude envelope introduced in
Sec.~\ref{sec:null_scalar_fraction}.

For the detector-level benchmark considered below, we take
\begin{equation}
    \mathcal M_c
    =
    8.49\,M_\odot,
    \qquad
    f_0
    =
    43.60\,{\rm mHz}.
    \label{eq:benchmark_parameters}
\end{equation}
The GW frequency reaches approximately
\begin{equation}
    f(T_{\rm obs})
    =
    53.00\,{\rm mHz}
    \label{eq:benchmark_final_frequency}
\end{equation}
after one year.

The benchmark is constructed using the detector-response setup of
Ref.~\cite{Xu:2026kev}.  In particular, the reference-sky calculation
retains one of the representative directions adopted there, and the
\((\mathcal M_c,f_0)\) plane is scanned at that fixed direction to
define the waveform used here.  Auxiliary scans at other representative
sky positions show that the precise conditional minimum shifts with
sky direction, while the corresponding scalar-fraction thresholds
remain at the few-\(10^{-3}\) level.  We therefore use
Eq.~\eqref{eq:benchmark_parameters} as a fixed benchmark rather than
as a sky-independent optimum.  The construction and the auxiliary
check are summarized in Appendix~\ref{app:benchmark_track}.

Once defined, the waveform parameters are held fixed in all-sky
and robustness calculations below.  Unless stated otherwise, we use
\begin{equation}
    T_{\rm obs}=1\,{\rm yr},
    \qquad
    \rho_T=1000,
    \qquad
    \rho_b^\star=5.
    \label{eq:reference_forecast_setup}
\end{equation}

\subsection{All-sky resolving power}
\label{subsec:allsky_reach}

We first evaluate the benchmark track on an equal-solid-angle sky
grid.  For the reference configuration in
Eq.~\eqref{eq:reference_forecast_setup}, the 10th, median, and
90th percentiles are
\begin{equation}
    \epsilon_{b,\min}^{10\%}
    =
    0.385405\%,
    \qquad
    \epsilon_{b,\min}^{50\%}
    =
    0.531921\%,
    \qquad
    \epsilon_{b,\min}^{90\%}
    =
    0.819262\%.
    \label{eq:allsky_main_percentiles}
\end{equation}
For this fixed benchmark track, direct sky refinement gives the most
favorable direction
\begin{equation}
    \beta_{\rm best}
    \simeq
    -70.98^\circ,
    \qquad
    \lambda_{\rm best}
    \simeq
    286.62^\circ,
    \label{eq:best_sky_direction}
\end{equation}
with
\begin{equation}
    \epsilon_{b,\min}^{\rm best}
    =
    0.306475\%.
    \label{eq:best_scalar_fraction}
\end{equation}

\begin{figure*}[t]
    \centering
    \includegraphics[
        width=0.52\textwidth
    ]{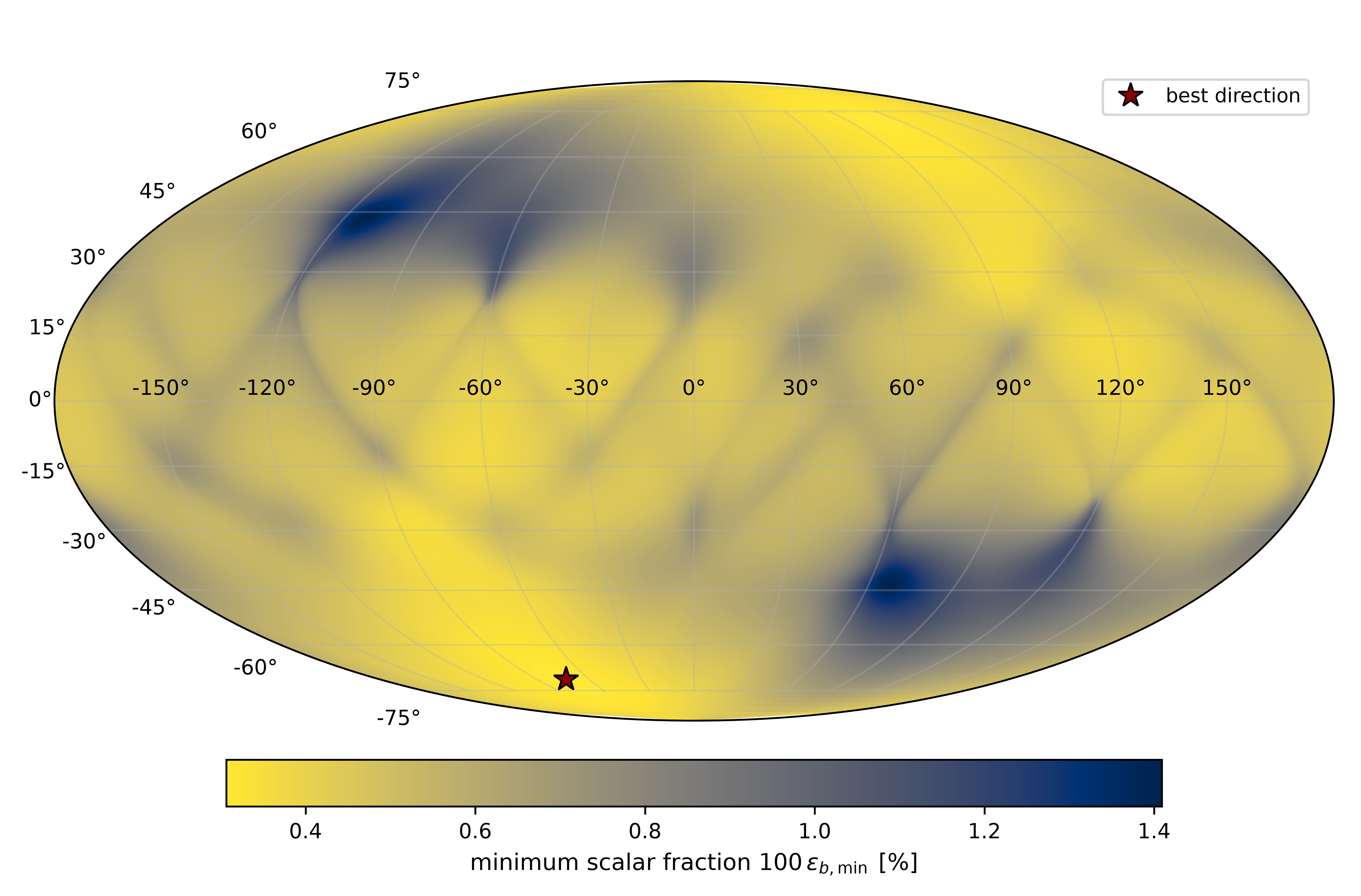}
    \hfill
    \includegraphics[
        width=0.43\textwidth
    ]{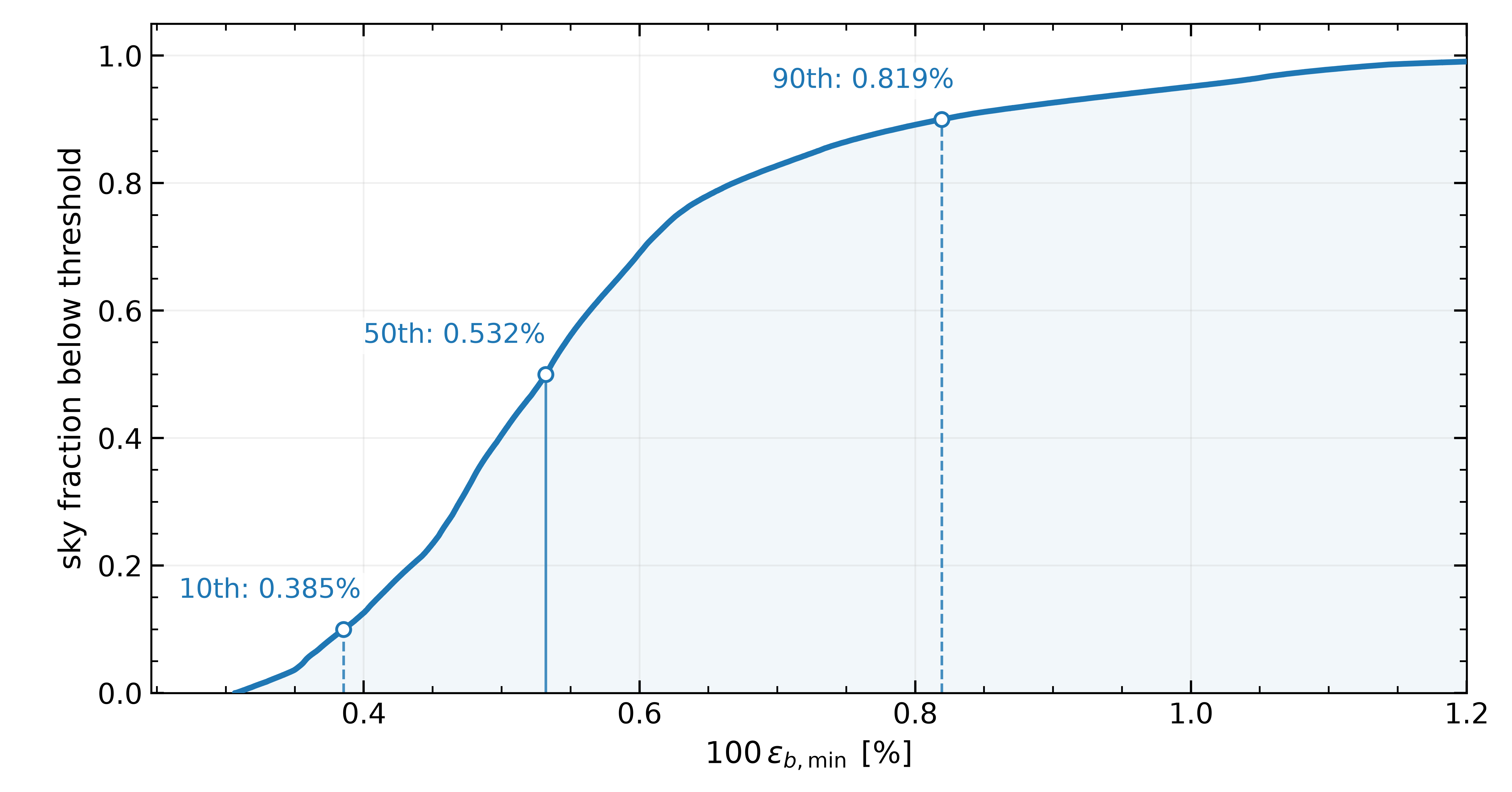}
    \caption{
    All-sky scalar-fraction resolving power for the benchmark chirp
    with \(\rho_T=1000\) and \(\rho_b^\star=5\).
    Left: equal-solid-angle sky map of the minimum resolvable
    transverse-scalar strain fraction; the star marks the most
    favorable direction for the fixed benchmark track.
    Right: cumulative sky fraction below a given scalar-fraction
    threshold.  The 10th, 50th, and 90th percentiles are
    \(0.385\%\), \(0.532\%\), and \(0.819\%\), respectively.
    }
    \label{fig:allsky_scalar_fraction}
\end{figure*}

The angular variation is shown in
Fig.~\ref{fig:allsky_scalar_fraction}.  At
\(\rho_T=1000\), \(40.7\%\) of the sky satisfies
\begin{equation}
    \epsilon_{b,\min}<0.5\%,
\end{equation}
while \(95.2\%\) satisfies
\begin{equation}
    \epsilon_{b,\min}<1\%.
\end{equation}
The median value in Eq.~\eqref{eq:allsky_main_percentiles} is used
below as the representative all-sky resolving power rather than the
minimum in Eq.~\eqref{eq:best_scalar_fraction}.

\subsection{Dependence on tensor SNR}
\label{subsec:rhoT_scaling}

For a fixed waveform track, sky position, and observation time,
changing the overall tensor amplitude changes \(\rho_T\) without
changing the detector information integrals \(I_T\) and \(I_b\).
Equation~\eqref{eq:epsilon_scaling} therefore gives
\begin{align}
    \epsilon_{b,\min}^{10\%}
    &\simeq
    0.3854\%
    \left(
        \frac{1000}{\rho_T}
    \right)
    \left(
        \frac{\rho_b^\star}{5}
    \right),
    \\
    \epsilon_{b,\min}^{50\%}
    &\simeq
    0.5319\%
    \left(
        \frac{1000}{\rho_T}
    \right)
    \left(
        \frac{\rho_b^\star}{5}
    \right),
    \label{eq:median_scalar_reach}
    \\
    \epsilon_{b,\min}^{90\%}
    &\simeq
    0.8193\%
    \left(
        \frac{1000}{\rho_T}
    \right)
    \left(
        \frac{\rho_b^\star}{5}
    \right).
\end{align}

\begin{figure}[t]
    \centering
    \includegraphics[width=0.55\linewidth]
    {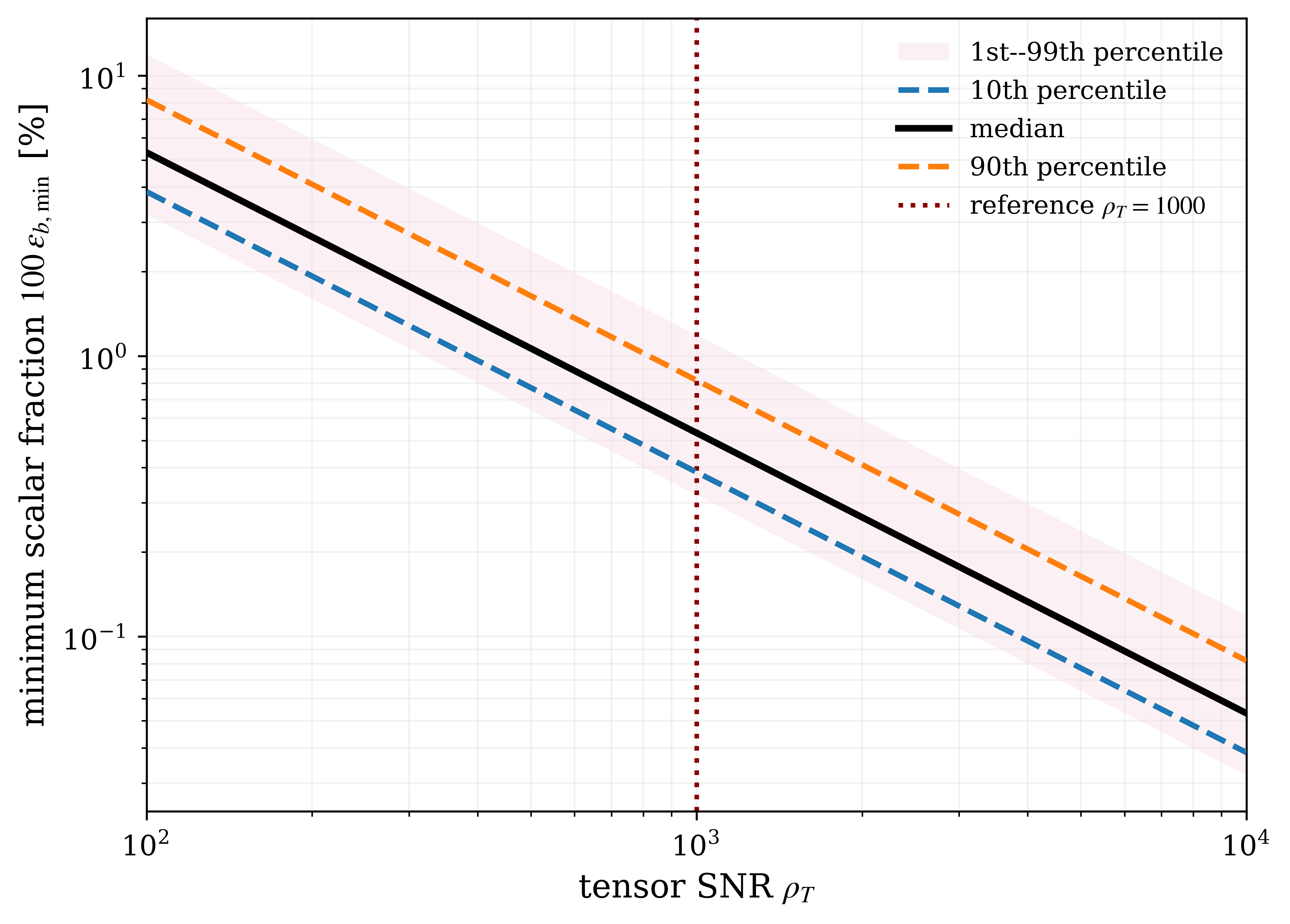}
    \caption{
    Minimum transverse-scalar strain fraction as a function of the
    tensor-network SNR for the benchmark chirp.  The solid curve gives
    the equal-area all-sky median, the dashed curves show the 10th and
    90th percentiles, and the shaded band spans the 1st--99th
    percentiles.  The vertical line marks the reference value
    \(\rho_T=1000\).  The scalar-channel threshold is
    \(\rho_b^\star=5\).
    }
    \label{fig:rhoT_scaling}
\end{figure}

Figure~\ref{fig:rhoT_scaling} shows this relation over
\(10^2\leq\rho_T\leq10^4\).
At the reference value \(\rho_T=1000\), the all-sky median threshold
is \(0.532\%\).  It decreases to \(0.266\%\) at
\(\rho_T=2000\) and to \(0.177\%\) at
\(\rho_T=3000\).

The corresponding sky coverage is summarized in
Table~\ref{tab:sky_coverage}.  Increasing the tensor SNR from
\(1000\) to \(2000\), for example, raises the fraction of the sky
reaching \(0.5\%\) from \(40.7\%\) to \(95.2\%\).  At
\(\rho_T=3000\), \(92.7\%\) of the sky reaches below
\(0.3\%\).

\begin{table}[t]
\caption{
Equal-area sky fraction satisfying three scalar-fraction thresholds
for the fixed benchmark waveform.  The scalar-channel threshold is
\(\rho_b^\star=5\).
}
\label{tab:sky_coverage}
\centering
\begin{ruledtabular}
\begin{tabular}{c c c c}
\(\rho_T\)
&
\(\epsilon_{b,\min}<0.3\%\)
&
\(\epsilon_{b,\min}<0.5\%\)
&
\(\epsilon_{b,\min}<1\%\)
\\
\hline
1000 & 0.0\%  & 40.7\%  & 95.2\%  \\
2000 & 69.1\% & 95.2\%  & 100.0\% \\
3000 & 92.7\% & 100.0\% & 100.0\%
\end{tabular}
\end{ruledtabular}
\end{table}

For the source-parameter robustness analysis below, we use the
equal-area median representative grid point
\begin{equation}
    \beta_{\rm med}
    =
    -15.0931^\circ,
    \qquad
    \lambda_{\rm med}
    =
    178.75^\circ,
    \label{eq:median_representative_sky}
\end{equation}
for which
\begin{equation}
    \epsilon_{b,\min}
    =
    0.531920\%.
    \label{eq:median_representative_fraction}
\end{equation}
This reproduces the all-sky median in
Eq.~\eqref{eq:allsky_main_percentiles} to the numerical precision
relevant here.

The results in this section assume an exactly tracked tensor signal.
Errors in the reconstructed sky position or chirp parameters break the
tensor-null conditions and produce residual tensor leakage.  We
quantify this effect in the next section.

\section{Robustness to source-parameter mismatch}
\label{sec:mismatch}

The scalar-fraction reach of the preceding section assumes that the
tensor-null combination is constructed from the correct source position
and chirp track.  In practice, these quantities must be inferred from the
tensor signal itself.  A mismatch in the assumed source parameters changes
the null direction and allows part of the tensor signal to enter the
nominally tensor-null channel.  This effect is potentially important for
a bright event: at the reference value \(\rho_T=1000\), a residual tensor
response at the few-\(10^{-3}\) level can already produce a leakage SNR
comparable to the adopted scalar threshold \(\rho_b^\star=5\).

We quantify this effect at the equal-area median representative direction
used in Sec.~\ref{sec:scalar_reach},
\begin{equation}
    \beta=-15.0931^\circ,
    \qquad
    \lambda=178.75^\circ,
\end{equation}
using the same benchmark chirp and A/E normalization as in the all-sky
calculation.

We parameterize the source mismatch by
\begin{equation}
    \Delta\boldsymbol{\theta}
    =
    \left(
        \Delta\beta,\,
        \Delta\lambda,\,
        \Delta\ln f_0,\,
        \Delta\ln\mathcal M_c,\,
        \Delta t_c/T_{\rm obs}
    \right).
    \label{eq:mismatch_coordinates}
\end{equation}

For the mismatch calculation, the chirp-track time coordinate
\(t_c\) enters through
\begin{equation}
    f(t;f_0,\mathcal M_c,t_c)
    =
    \left[
        f_0^{-8/3}
        -
        \frac{256}{5}\pi^{8/3}
        \left(
            \frac{G\mathcal M_c}{c^3}
        \right)^{5/3}
        (t-t_c)
    \right]^{-3/8}.
    \label{eq:mismatch_chirp_track}
\end{equation}
The true benchmark corresponds to \(t_c=0\).  Here \(t_c\) is a
time-origin coordinate of the chirp track and should not be interpreted
as an independent arrival-time offset of the rapidly varying carrier.

\subsection{One-dimensional tracking errors}
\label{subsec:one_dimensional_mismatch}

We first vary the five coordinates in
Eq.~\eqref{eq:mismatch_coordinates} one at a time.  At each displaced
point the tensor-null combination is reconstructed from the assumed
parameters and then applied to the tensor signal generated with the true
parameters.  We evaluate both the polarization-averaged tensor leakage
and the largest leakage over tensor polarization states.

\begin{figure}[t]
    \centering
    \includegraphics[width=\linewidth]
    {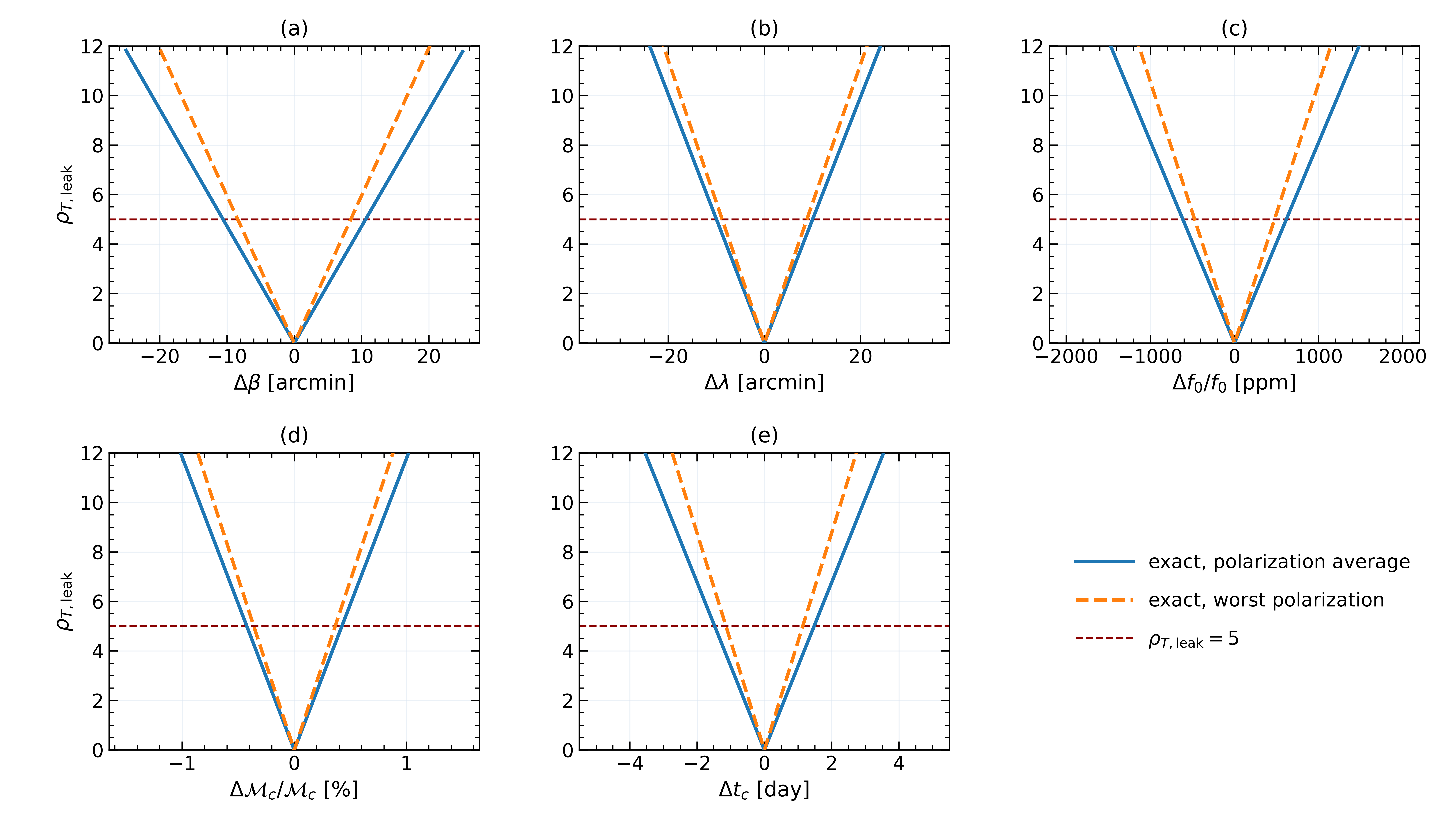}
    \caption{
    Tensor leakage produced by one-parameter mismatch at the
    equal-area median representative sky position.
    The curves show the exact polarization-averaged and
    worst-polarization leakage SNR for errors in sky position,
    initial frequency, chirp mass, and the chirp-track time coordinate.
    The horizontal reference level is
    \(\rho_{T,\mathrm{leak}}=5\).
    }
    \label{fig:one_dimensional_mismatch}
\end{figure}

Figure~\ref{fig:one_dimensional_mismatch} shows the resulting exact
leakage SNRs.  As expected, the residual vanishes at zero mismatch and
grows rapidly once the assumed sky position or chirp track departs from
the true one.  We use
\begin{equation}
    \rho_{T,\mathrm{leak}}^{\rm worst}=5
    \label{eq:leakage_reference_level}
\end{equation}
as a conservative reference level, equal to the scalar SNR adopted in the
resolving-power forecast.  Taking the smaller absolute crossing on the
two sides of zero gives
\begin{equation}
\begin{array}{c|c}
    \text{mismatched parameter}
    &
    \rho_{T,\mathrm{leak}}^{\rm worst}=5
    \\ \hline
    \beta
    &
    8.39086~{\rm arcmin}
    \\
    \lambda
    &
    8.78456~{\rm arcmin}
    \\
    f_0
    &
    474.31~{\rm ppm}
    \\
    \mathcal M_c
    &
    0.360855\%
    \\
    t_c
    &
    1.1392~{\rm day}
\end{array}
\label{tab:one_dimensional_mismatch}
\end{equation}

These numbers are not parameter-estimation uncertainties. They give the
one-at-a-time displacement at which an incorrectly constructed null
channel admits tensor contamination at the same SNR scale as the scalar
signal considered here. In particular, they should not be interpreted
as five independent cuts on an allowed source-parameter region.
\footnote{For comparison, sub-(10)-arcmin sky localization has been
reported for favorable massive-black-hole binaries in LISA
parameter-estimation studies \cite{McWilliams:2011zs}. Whether the
benchmark considered here can achieve the required tracking accuracy
would require a dedicated parameter-estimation analysis.}

\subsection{Joint mismatch geometry}
\label{subsec:joint_mismatch}

The latter distinction matters because the five tracking coordinates are
correlated.  Near the true parameters, the polarization-averaged leakage
is well described by a quadratic form,
\begin{equation}
    \left(
        \rho_{T,\mathrm{leak}}^{\rm avg}
    \right)^2
    =
    \Delta\boldsymbol{\theta}^{\,T}
    G_{\rm leak}
    \Delta\boldsymbol{\theta}
    +
    O(\Delta\theta^3),
    \label{eq:joint_leakage_metric}
\end{equation}
where \(G_{\rm leak}\) is positive semidefinite.  To compare parameters
with different units, we define
\begin{equation}
    u_i=\frac{\Delta\theta_i}{s_i},
\end{equation}
where \(s_i\) is the local one-dimensional scale for which the
polarization-averaged leakage reaches 5.
Equation
\eqref{eq:joint_leakage_metric} can then be written as
\begin{equation}
    \left(
        \rho_{T,\mathrm{leak}}^{\rm avg}
    \right)^2
    =
    25\,\mathbf u^{T}C\mathbf u,
    \qquad
    C_{ii}=1.
    \label{eq:normalized_leakage_metric}
\end{equation}

\begin{figure}[t]
    \centering
    \includegraphics[width=0.75\linewidth]
    {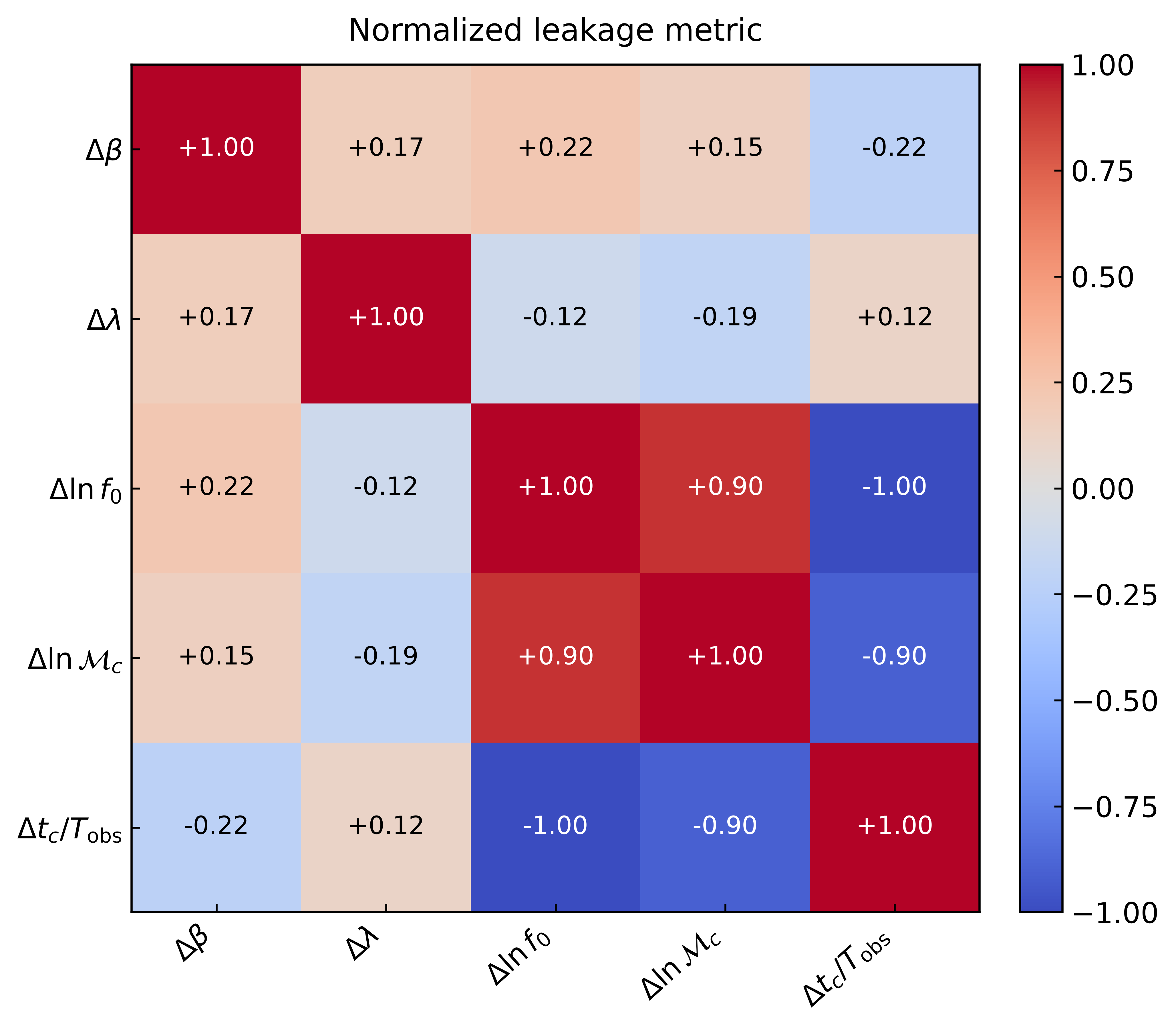}
    \caption{
    Normalized polarization-averaged tensor-leakage metric \(C\)
    for the five source-tracking coordinates in
    Eq.~\eqref{eq:mismatch_coordinates}.
    The intrinsic chirp parameters form a strongly correlated block,
    with an almost exact degeneracy between
    \(\ln f_0\) and \(t_c/T_{\rm obs}\), while their coupling to the
    two sky coordinates is comparatively weak.
    }
    \label{fig:joint_leakage_metric}
\end{figure}

The normalized matrix \(C\) is shown in
Fig.~\ref{fig:joint_leakage_metric}.  Its eigenvalues are
\begin{equation}
    2.9523,\quad
    1.1708,\quad
    0.7523,\quad
    0.1247,\quad
    3.2\times10^{-14},
    \label{eq:leakage_metric_eigenvalues}
\end{equation}
so that only four independent leakage directions are locally resolved,
\begin{equation}
    \operatorname{rank}G_{\rm leak}=4.
    \label{eq:leakage_metric_rank}
\end{equation}
The nearly null eigenvector is, to a very good approximation,
\begin{equation}
    \mathbf v_{\rm null}
    \simeq
    \frac{1}{\sqrt{2}}
    \left(
        0,\,
        0,\,
        1,\,
        0,\,
        1
    \right),
    \label{eq:leakage_null_direction}
\end{equation}
corresponding to a joint displacement of
\(\ln f_0\) and \(t_c/T_{\rm obs}\).

This degeneracy follows directly from
Eq.~\eqref{eq:mismatch_chirp_track}.  In terms of
\(f^{-8/3}\), both \(f_0\) and \(t_c\) modify the constant,
or time-origin, part of the frequency track.  An appropriate correlated
displacement can therefore leave the local track nearly unchanged.
The same matrix
also shows substantial correlations between
\(\ln f_0\), \(\ln\mathcal M_c\), and \(t_c/T_{\rm obs}\), whereas the
coupling between the sky coordinates and the intrinsic chirp parameters
is much weaker.  The allowed local mismatch region is therefore strongly
anisotropic rather than a rectangular product of the five
one-dimensional intervals.

The metric in Eq.~\eqref{eq:joint_leakage_metric} refers to the
polarization-averaged leakage.  For the conservative worst-polarization
quantity we retain the full \(2\times2\) tensor-polarization leakage
matrix and determine its largest generalized eigenvalue; a single
ordinary \(5\times5\) matrix does not represent that maximization for
all mismatch directions.  Exact recomputations along the resolved
principal directions confirm the local quadratic description in the
region relevant to \(\rho_{T,\mathrm{leak}}\lesssim5\).  The numerical
tests and the exact-versus-local comparisons are given in
Appendix~\ref{app:leakage_validation}.

The sub-percent reach obtained in Sec.~\ref{sec:scalar_reach} is
therefore conditional not only on a bright tensor signal, but also on
sufficiently accurate tracking of its sky position and frequency
evolution.  The mismatch calculation translates this condition into a
direct tensor-leakage requirement.  It does not determine how frequently
that requirement will be satisfied by parameter estimation for an
astrophysical source; doing so would require combining the leakage
geometry with a Fisher or Bayesian posterior for the tensor signal.

\section{Discussion and conclusions}
\label{sec:discussion}

We have studied the detector-level resolving power for a
co-propagating transverse-scalar polarization in an already identified
bright tensor chirp.  The calculation assumes that the sky position
and tensor time-frequency track are known when constructing the
source-tracked tensor-null response, while the scalar amplitude is
left free.  This separates the question of polarization resolvability
from the model-dependent question of how large a scalar component is
produced by a particular source.

The transverse-scalar mode considered here is motivated by GQFT, whose
linearized gravitational dynamics contains an additional scalar
polarization besides the two tensor modes of GR
\cite{Gao:2025aye}.  Our detector construction builds on the dynamic
null-response formalism of Ref.~\cite{Xu:2026kev}, but differs from a
monochromatic response study in that the tensor-null coefficients are
tracked along an evolving chirp and the scalar fraction is normalized
to the unequal-arm \(A/E\) tensor network.

For the one-year benchmark chirp used throughout this work, with
\(\rho_T=1000\) and an operational scalar-channel threshold
\(\rho_b^\star=5\), the equal-solid-angle sky distribution gives
\begin{equation}
    \epsilon_{b,\min}^{10\%}
    =
    0.3854\%,
    \qquad
    \epsilon_{b,\min}^{50\%}
    =
    0.5319\%,
    \qquad
    \epsilon_{b,\min}^{90\%}
    =
    0.8193\%.
\end{equation}
The most favorable direction for this fixed benchmark track reaches
\begin{equation}
    \epsilon_{b,\min}^{\rm best}
    =
    0.3065\%.
\end{equation}
The median rather than the best-sky value provides a more useful
summary of the detector-level reach.  At the reference tensor SNR,
\(95.2\%\) of the sky can probe a scalar fraction below \(1\%\), while
\(40.7\%\) reaches below \(0.5\%\).

The scalar-fraction threshold scales inversely with the tensor SNR,
\begin{equation}
    \epsilon_{b,\min}
    \propto
    \frac{\rho_b^\star}{\rho_T},
\end{equation}
so that the median result can be written as
\begin{equation}
    \epsilon_{b,\min}^{50\%}
    \simeq
    0.5319\%
    \left(
        \frac{1000}{\rho_T}
    \right)
    \left(
        \frac{\rho_b^\star}{5}
    \right).
    \label{eq:discussion_median_scaling}
\end{equation}
The reference choice \(\rho_T=1000\) should therefore be regarded as a
convenient normalization for a bright event rather than as a special
SNR scale.

The waveform used in the forecast is likewise a benchmark rather than
a global optimum.  Its parameters were selected at a reference sky
direction inherited from the detector-response setup of
Ref.~\cite{Xu:2026kev}.  Repeating the conditional
\((\mathcal M_c,f_0)\) scan at the median representative sky and at the
most favorable benchmark-track direction changes the location of the
minimum.  Nevertheless, the three tests remain within the same broad
waveform regime: the preferred chirp masses are of order a few solar
masses, the initial frequencies lie in the \(40\)--\(50\,{\rm mHz}\)
band, and the optimized scalar-fraction thresholds remain at the
few-\(10^{-3}\) level.  This supports the use of the chosen track as a
detector benchmark, but not its interpretation as a sky-independent
optimal waveform.

The ideal resolving power also requires sufficiently accurate
knowledge of the tensor source parameters.  At the equal-area median
representative sky, one-parameter mismatches produce
worst-polarization tensor leakage with
\(\rho_{T,\mathrm{leak}}=5\) at approximately
\begin{equation}
    |\Delta\beta|
    \simeq
    8.39~{\rm arcmin},
    \qquad
    |\Delta\lambda|
    \simeq
    8.78~{\rm arcmin},
\end{equation}
\begin{equation}
    \frac{|\Delta f_0|}{f_0}
    \simeq
    4.74\times10^{-4},
    \qquad
    \frac{|\Delta\mathcal M_c|}{\mathcal M_c}
    \simeq
    3.61\times10^{-3},
\end{equation}
and
\begin{equation}
    |\Delta t_c|
    \simeq
    1.14~{\rm day}.
\end{equation}
These values characterize leakage tolerances and are not
parameter-estimation uncertainties.

The joint mismatch calculation further shows that the five nominal
tracking coordinates contain only four locally resolved leakage
directions.  The nearly null direction is dominated by a correlated
shift of \(\ln f_0\) and \(t_c/T_{\rm obs}\), reflecting the local
redundancy of these quantities in the leading-order chirp
parameterization.  Strong correlations are also present among
\(f_0\), \(\mathcal M_c\), and \(t_c\), whereas their coupling to the
sky coordinates is comparatively weak.  The relevant tracking
requirement is therefore a correlated region in parameter space rather
than the Cartesian product of five independent one-dimensional
bounds.

Several limitations of the present forecast follow directly from its
scope.  The benchmark track is described by a leading-order
quasi-circular inspiral and is used to characterize detector response,
not to identify an astrophysically preferred source population.
Likewise, the local leakage metric propagates a specified source
mismatch into tensor contamination, but does not determine the
posterior uncertainties with which the tensor source parameters can be
measured.  Establishing whether a given event satisfies the tracking
requirements obtained here requires a parameter-estimation analysis
using a consistent tensor waveform model.

The sky dependence found in the benchmark checks also leaves a
well-defined optimization problem.  Rather than minimizing
\(\epsilon_{b,\min}\) at one fixed sky position, one may optimize
waveform parameters according to quantities such as
\begin{equation}
    \operatorname*{median}_{\Omega}
    \epsilon_{b,\min},
\end{equation}
other sky percentiles, or a source-population-weighted statistic.
The present auxiliary scans already indicate that different sky
directions favor different chirp tracks while retaining a similar
overall sensitivity scale.  A systematic study of this waveform--sky
optimization is therefore naturally separate from the fixed-benchmark
forecast considered here.

A second step is to connect the detector-level scalar-fraction reach to
specific source models.  In GQFT, the scalar amplitude is determined by
the source dynamics rather than being an independent phenomenological
parameter \cite{Gao:2025aye}.  Determining which astrophysical systems
can produce both a sufficiently bright tensor signal and a scalar
fraction in the range probed here requires source-dependent waveform
generation and, ultimately, a population calculation.  That question
is deliberately left open in the present analysis.

The main result can therefore be stated conditionally but without
reference to a particular source model: for a bright and accurately
tracked tensor chirp with \(\rho_T\sim10^3\), the dynamic tensor-null
response can probe a co-propagating transverse-scalar strain fraction
at the few-\(10^{-3}\) level over a substantial fraction of the sky.
The achievable threshold improves linearly with tensor SNR, while the
mismatch analysis specifies the source-tracking accuracy required for
the tensor leakage to remain below the same scale.

\begin{acknowledgments}
The authors acknowledge HIAS for access to the ``Quantum Universe
Physical Simulation Platform''. B.-X. G. acknowledges support from
the National Natural Science Foundation of China (NSFC) under Grant
No.~12505066. Z. C. acknowledges support from the National Key
Research and Development Program of China under Grant
No.~2021YFC2203001 and the National Natural Science Foundation of
China (NSFC) under Grant No.~12475049.

\end{acknowledgments}

\clearpage
\appendix

\section{Selection of the reference chirp}
\label{app:benchmark_track}

The calculations in Sec.~\ref{sec:scalar_reach} require a fixed
time-frequency track.  We construct this benchmark using the
detector-response setup of Ref.~\cite{Xu:2026kev}.  In particular,
we retain the first of the representative sky directions used there,
\begin{equation}
    \beta_{\rm ref}
    =
    0.88,
    \qquad
    \lambda_{\rm ref}
    =
    1.77,
    \label{eq:appendix_reference_sky}
\end{equation}
in radians.  In the sky convention used here, these correspond to an
ecliptic latitude and longitude of approximately
\(50.42^\circ\) and \(101.41^\circ\), respectively.  The direction is
used only to define the benchmark waveform and is not assumed to be an
optimal sky position.

At this fixed direction, we evaluate the scalar-fraction statistic
introduced in Sec.~\ref{sec:null_scalar_fraction},
\begin{equation}
    \epsilon_{b,\min}
    =
    \frac{\rho_b^\star}{\rho_T}
    \sqrt{\frac{I_T}{I_b}},
    \label{eq:appendix_scalar_fraction_statistic}
\end{equation}
over the \((\mathcal M_c,f_0)\) plane.  The observation time and SNR
normalization are kept at
\begin{equation}
    T_{\rm obs}=1\,{\rm yr},
    \qquad
    \rho_T=1000,
    \qquad
    \rho_b^\star=5.
    \label{eq:appendix_benchmark_setup}
\end{equation}
Thus, the sky position and detector configuration are fixed in this
scan, while \(\mathcal M_c\) and \(f_0\) determine the one-year chirp
track and hence the accumulated information integrals \(I_T\) and
\(I_b\).

\begin{figure*}[t]
    \centering
    \includegraphics[
        width=0.88\textwidth
    ]{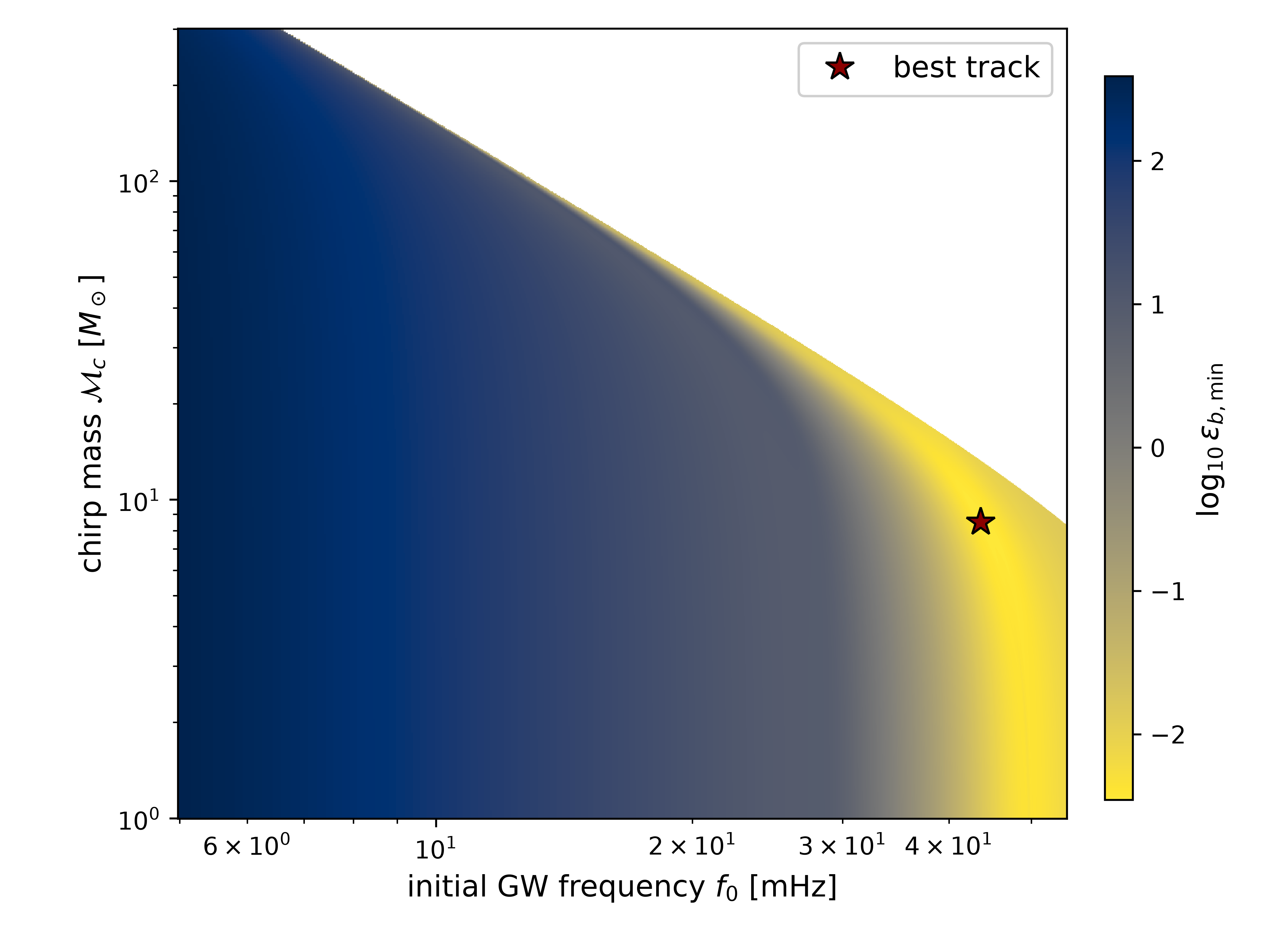}
    \caption{
    Detector-level scan used to construct the benchmark chirp.
    The minimum resolvable transverse-scalar fraction is evaluated in
    the \((\mathcal M_c,f_0)\) plane at the fixed reference direction
    \((\beta_{\rm ref},\lambda_{\rm ref})=(0.88,1.77)\).
    The marker denotes the refined point adopted as the benchmark
    waveform in the main analysis.  The scan is conditional on this
    reference direction and is not a joint optimization over waveform
    and sky parameters.
    }
    \label{fig:appendix_chirp_map}
\end{figure*}

Figure~\ref{fig:appendix_chirp_map} shows the resulting detector-level
scan.  A broad survey is followed by a high-resolution refinement of
the favorable region.  The track retained for the main analysis is
\begin{equation}
    \mathcal M_c
    =
    8.490919336\,M_\odot,
    \qquad
    f_0
    =
    43.598750\,{\rm mHz},
    \label{eq:appendix_exact_benchmark}
\end{equation}
which evolves to
\begin{equation}
    f(T_{\rm obs})
    =
    52.999506\,{\rm mHz}
    \label{eq:appendix_final_frequency}
\end{equation}
after one year.  At the reference direction in
Eq.~\eqref{eq:appendix_reference_sky}, this track gives
\begin{equation}
    \epsilon_{b,\min}
    =
    0.349136\%.
    \label{eq:appendix_reference_threshold}
\end{equation}

The refined point is well separated from the boundaries of the scanned
domain.  A direct recomputation of the full chirp gives a relative
difference of
\begin{equation}
    4.81\times10^{-7}
\end{equation}
from the interpolated detector-kernel result.  The fraction of the
one-year track retained after imposing the t-NRC conditioning cut is
\begin{equation}
    0.995893.
\end{equation}
These checks show that the location of the selected point in
Fig.~\ref{fig:appendix_chirp_map} is not set by interpolation error,
conditioning loss, or a scan boundary.

The scan in Fig.~\ref{fig:appendix_chirp_map} is nevertheless
conditional on the reference sky.  To assess how strongly the
selection depends on this choice, we repeated the
\((\mathcal M_c,f_0)\) scan at two additional directions: the
equal-area median representative sky used in the robustness analysis
and the most favorable direction found for the fixed benchmark track
in Sec.~\ref{subsec:allsky_reach}.  The resulting conditional minima
are summarized in Table~\ref{tab:appendix_reference_sky_check}.

\begin{table*}[t]
\caption{
Conditional minima of the \((\mathcal M_c,f_0)\) scan at three sky
directions.  The first row is the reference direction inherited from
Ref.~\cite{Xu:2026kev}; the other two are obtained from the independent
all-sky calculation for the benchmark track.  All entries use
\(T_{\rm obs}=1\,{\rm yr}\), \(\rho_T=1000\), and
\(\rho_b^\star=5\).
}
\label{tab:appendix_reference_sky_check}
\centering
\begin{ruledtabular}
\begin{tabular}{l c c c c c}
sky
&
\(\beta\,[{\rm deg}]\)
&
\(\lambda\,[{\rm deg}]\)
&
\(\mathcal M_c\,[M_\odot]\)
&
\(f_0\,[{\rm mHz}]\)
&
\(\epsilon_{b,\min}\)
\\
\hline
reference direction
&
50.4203
&
101.4135
&
8.4909
&
43.5988
&
0.3491\%
\\
median representative sky
&
-15.0931
&
178.7500
&
2.8603
&
49.5571
&
0.3233\%
\\
most favorable benchmark-track sky
&
-70.9803
&
286.6170
&
6.7423
&
45.2907
&
0.2979\%
\end{tabular}
\end{ruledtabular}
\end{table*}

The preferred waveform parameters are therefore not independent of sky
position.  The variation is particularly pronounced in chirp mass,
while the preferred initial frequencies in these tests remain within
the same tens-of-millihertz band.  At the same time, the three
conditional minima all remain at the few-\(10^{-3}\) level.  The
benchmark in Eq.~\eqref{eq:appendix_exact_benchmark} therefore captures
the characteristic scale of the detector-level resolving power found
in these tests, but it should not be interpreted as a sky-independent
optimal waveform.

For this reason, the waveform parameters in
Eq.~\eqref{eq:appendix_exact_benchmark} are held fixed after the
benchmark is defined.  The all-sky results in
Sec.~\ref{subsec:allsky_reach} then characterize the angular dependence
of this single waveform rather than reoptimizing the source parameters
at each sky position.  A systematic search for waveform parameters
optimized over sky-averaged, percentile-based, or population-weighted
criteria constitutes a separate problem and is left for future work.


\section{Frequency and observation-time dependence}
\label{app:frequency_time}

The main analysis integrates the detector information along an evolving
chirp.  It is also useful to examine the underlying frequency dependence
before performing this integration.  We therefore consider a
monochromatic transverse-scalar signal at fixed GW frequency \(f\) and
evaluate the annually averaged detector information 
using the same orbital response, t-NRC construction, conditioning
criterion, and unequal-arm \(A/E\) tensor normalization as in the main
calculation.

For a monochromatic breathing strain \(h_b\), define the time-averaged
information rate
\begin{equation}
    \overline{K}_b(f)
    =
    \frac{1}{T_{\rm yr}}
    \int_0^{T_{\rm yr}}
    K_b(f,t)\,dt ,
    \label{eq:appendix_kb_average}
\end{equation}
where epochs failing the t-NRC conditioning requirement are assigned
zero weight.  Using this annual mean as the effective information rate,
the strain required to reach a scalar-channel SNR
\(\rho_b^\star\) over an observation time \(T_{\rm obs}\) is
\begin{equation}
    h_{b,\min}(f,T_{\rm obs})
    =
    \frac{\rho_b^\star}
    {\sqrt{T_{\rm obs}\,\overline{K}_b(f)}} .
    \label{eq:appendix_hb_requirement}
\end{equation}
For the one-year result below, Eq.~\eqref{eq:appendix_hb_requirement}
uses the information rate obtained directly from the full annual
detector calculation.  Its extension to longer observing times is
specified below.

Similarly, with
\(\overline{K}_T(f)\) denoting the annually averaged unequal-arm
\(A/E\) tensor information rate, the corresponding scalar-fraction
requirement at a fixed tensor-network SNR is
\begin{equation}
    \epsilon_{b,\min}(f)
    =
    \frac{\rho_b^\star}{\rho_T}
    \sqrt{
        \frac{\overline{K}_T(f)}
             {\overline{K}_b(f)}
    }.
    \label{eq:appendix_monochromatic_fraction}
\end{equation}
Equation~\eqref{eq:appendix_monochromatic_fraction} is the
monochromatic counterpart of the chirp-integrated statistic used in
Sec.~\ref{sec:scalar_reach}.

\begin{figure*}[t]
    \centering
    \includegraphics[
        width=0.92\textwidth
    ]{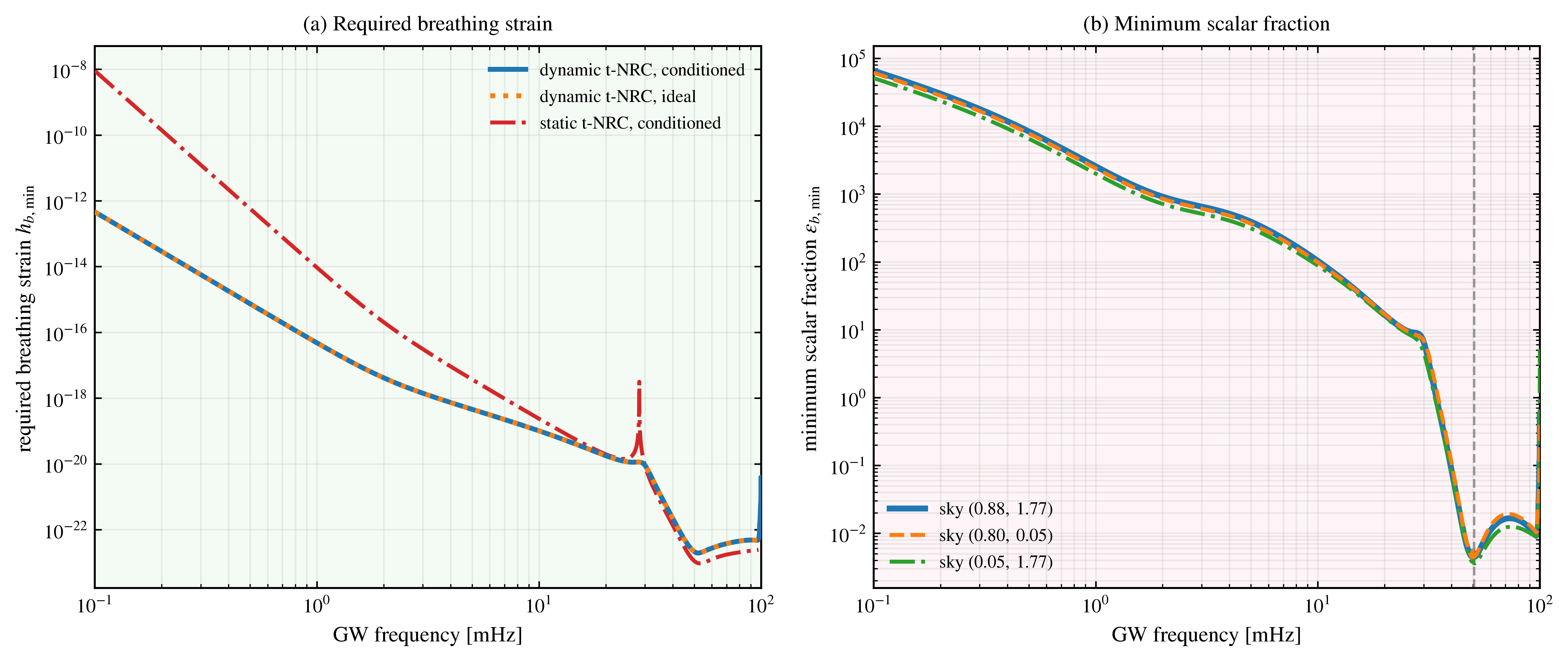}
    \caption{
    Frequency dependence of the transverse-scalar forecast for a
    one-year observation.
    Left: breathing strain required to reach
    \(\rho_b^\star=5\) at the primary reference sky.  The dynamic
    conditioned t-NRC is compared with the ideal dynamic construction
    and with a conditioned static constellation.
    Right: minimum scalar fraction for
    \(\rho_T=1000\) and \(\rho_b^\star=5\) at three representative
    sky directions, using the dynamic conditioned t-NRC and the
    unequal-arm \(A/E\) tensor network.
    The favorable response lies in the tens-of-millihertz band.
    }
    \label{fig:appendix_frequency_forecast}
\end{figure*}

Figure~\ref{fig:appendix_frequency_forecast} shows the frequency
dependence of these quantities.  The left panel compares the absolute
breathing-strain requirement for the dynamic conditioned t-NRC with
the corresponding ideal dynamic and conditioned static constructions.
For the static comparison, we use the fixed equilateral constellation
geometry adopted in Ref.~\cite{Xu:2026kev}, while keeping the same
source direction and t-NRC conditioning criterion.
The right panel shows the scalar-fraction threshold for three
representative sky directions using the dynamic conditioned t-NRC and
the unequal-arm \(A/E\) tensor network.

The frequency curves show a broad favorable region at a few times
\(10^{-2}\,\mathrm{Hz}\), rather than a narrow isolated feature.
For the primary sky, the minimum absolute breathing-strain requirement
occurs at
\begin{equation}
    f
    =
    50.6633\,{\rm mHz},
    \label{eq:appendix_best_monochromatic_frequency}
\end{equation}
where a one-year observation with
\(\rho_b^\star=5\) requires
\begin{equation}
    h_{b,\min}
    =
    2.03684\times10^{-23}.
    \label{eq:appendix_best_monochromatic_strain}
\end{equation}
The scalar-fraction curves at the other sky directions have their
minima in the same broad frequency range.  The benchmark chirp used in
the main analysis,
\begin{equation}
    43.60\,{\rm mHz}
    \lesssim
    f(t)
    \lesssim
    53.00\,{\rm mHz},
\end{equation}
therefore evolves through the same detector-response window identified
by the monochromatic calculation.  This agreement is only a
consistency check: the benchmark itself is defined by the
two-dimensional chirp scan described in
Appendix~\ref{app:benchmark_track}, rather than by the monochromatic
minimum of Eq.~\eqref{eq:appendix_best_monochromatic_frequency}.

The comparison between the ideal and conditioned dynamic curves also
shows that the favorable frequency window is retained after excluding
poorly conditioned tensor-null epochs.  Its location is therefore not
set by a numerical singularity of the t-NRC construction.  The
difference between the dynamic and static curves reflects the annual
orbital modulation of the constellation, which is included throughout
the main chirp calculation.

For observations extending beyond one year, we assume that the
annually averaged information rate remains unchanged and repeats from
year to year.  Under this assumption,
Eq.~\eqref{eq:appendix_hb_requirement} gives
\begin{equation}
    h_{b,\min}
    \propto
    T_{\rm obs}^{-1/2},
    \qquad
    T_{\rm obs}\geq1\,{\rm yr}.
    \label{eq:appendix_time_scaling}
\end{equation}
At the frequency in
Eq.~\eqref{eq:appendix_best_monochromatic_frequency}, the corresponding
values are listed in Table~\ref{tab:appendix_time_scaling}.


\begin{table}[t]
\caption{
Observation-time dependence of the required monochromatic breathing
strain at \(f=50.6633\,{\rm mHz}\) and
\(\rho_b^\star=5\).  The one-year value is obtained from the full
annual detector calculation; longer observations assume that the
annually averaged information rate repeats from year to year.
}
\label{tab:appendix_time_scaling}
\centering

\begin{minipage}{0.55\columnwidth}
\begin{ruledtabular}
\begin{tabular}{c c}
\(T_{\rm obs}\,[{\rm yr}]\)
&
\(h_{b,\min}\)
\\
\hline
1.00 & \(2.03684\times10^{-23}\) \\
2.00 & \(1.44027\times10^{-23}\) \\
4.00 & \(1.01842\times10^{-23}\)
\end{tabular}
\end{ruledtabular}
\end{minipage}

\end{table}


These values are accurately described by
\begin{equation}
    h_{b,\min}
    \simeq
    2.03684\times10^{-23}
    \left(
        \frac{\rho_b^\star}{5}
    \right)
    \left(
        \frac{T_{\rm obs}}{1\,{\rm yr}}
    \right)^{-1/2},
    \qquad
    T_{\rm obs}\geq1\,{\rm yr},
    \label{eq:appendix_time_scaling_fit}
\end{equation}
at the frequency and sky position considered here.

This time scaling applies to the absolute strain requirement.
When the tensor signal is instead normalized to a fixed observed
network SNR \(\rho_T\), both the tensor and scalar strain scales acquire
the same \(T_{\rm obs}^{-1/2}\) factor under the same annual-rate
assumption.  Their ratio therefore reduces to
Eq.~\eqref{eq:appendix_monochromatic_fraction}, and the
scalar-fraction forecast is controlled by the relative tensor and
breathing information rates rather than by the observation time alone.

The monochromatic calculation thus provides two checks on the
chirp-resolved forecast.  The frequency interval traversed by the
benchmark lies within the broad favorable response band of the dynamic
detector, and, for observations extending beyond one year, the
absolute strain requirement follows the expected
coherent-information scaling under the repeated annual-rate
approximation.

\section{Validation of the local tensor-leakage metric}
\label{app:leakage_validation}

The source-tracking analysis in the main text uses a local quadratic
description of the tensor leakage induced by errors in the assumed
source parameters.  Here we compare that approximation directly with
the full mismatched tensor-null calculation and summarize the numerical
checks used in constructing the joint leakage metric.

We use the five mismatch coordinates
\begin{equation}
    \boldsymbol{\theta}
    =
    \left(
        \beta,\,
        \lambda,\,
        \ln f_0,\,
        \ln\mathcal M_c,\,
        t_c/T_{\rm obs}
    \right),
    \label{eq:appendix_mismatch_coordinates}
\end{equation}
and denote the true source parameters by
\(\boldsymbol{\theta}_0\).
The chirp-track time coordinate \(t_c\) is defined by
Eq.~\eqref{eq:mismatch_chirp_track}.

For an assumed parameter offset
\(\Delta\boldsymbol{\theta}\), the tensor-null coefficient is
constructed from
\(\boldsymbol{\theta}_0+\Delta\boldsymbol{\theta}\) and applied to the
true tensor response.  The residual response for
\(p=+,\times\) is
\begin{equation}
    \ell_p
    \left(
        t;\Delta\boldsymbol{\theta}
    \right)
    =
    a_t^I
    \left(
        t;\boldsymbol{\theta}_0+
        \Delta\boldsymbol{\theta}
    \right)
    R_{p,I}^{\rm Sag}
    \left(
        t;\boldsymbol{\theta}_0
    \right),
    \label{eq:appendix_leakage_amplitude}
\end{equation}
where repeated Sagnac-channel indices are summed.  At zero mismatch,
\(\ell_+=\ell_\times=0\) up to numerical roundoff.

The noise power of the same mismatched tensor-null combination is
\begin{equation}
\begin{aligned}
    S_{\Delta\boldsymbol{\theta}}(t)
    ={}&
    a_t^I
    \left(
        t;\boldsymbol{\theta}_0+
        \Delta\boldsymbol{\theta}
    \right)
    \left(S_N^{\rm Sag}\right)_{IJ}
    \left(
        t;\boldsymbol{\theta}_0
    \right)
    a_t^{J*}
    \left(
        t;\boldsymbol{\theta}_0+
        \Delta\boldsymbol{\theta}
    \right),
    \\
    S_0(t)
    \equiv{}&
    S_{\Delta\boldsymbol{\theta}=0}(t).
\end{aligned}
\label{eq:appendix_mismatched_noise}
\end{equation}
Thus \(S_0(t)\) is the projected tensor-null noise power at the true
source parameters, whereas
\(S_{\Delta\boldsymbol{\theta}}(t)\) is recomputed with the mismatched
null coefficient.

For sufficiently small offsets we expand
\begin{equation}
    \ell_p
    =
    \sum_i
    D_{pi}(t)\,
    \Delta\theta_i
    +
    O(\Delta\theta^2),
    \label{eq:appendix_leakage_derivative}
\end{equation}
where
\begin{equation}
    D_{pi}(t)
    =
    \left.
    \frac{\partial\ell_p}
         {\partial\theta_i}
    \right|_{\boldsymbol{\theta}_0}.
\end{equation}
The derivatives are evaluated numerically with symmetric finite
differences and Richardson extrapolation.

Let \(I_T\) be the polarization-averaged tensor information of the
unequal-arm \(A/E\) network defined in the main text.
The polarization-averaged leakage SNR then has the local form
\begin{equation}
    \rho_{T,{\rm leak,avg}}^2
    =
    \Delta\boldsymbol{\theta}^{T}
    \mathbf G_{\rm leak}
    \Delta\boldsymbol{\theta}
    +
    O(\Delta\theta^3),
    \label{eq:appendix_local_metric}
\end{equation}
with
\begin{equation}
    \left(G_{\rm leak}\right)_{ij}
    =
    \frac{\rho_T^2}{2I_T}
    \int dt\,
    \frac{\mathcal A^2(t)}{S_0(t)}
    {\rm Re}
    \left[
        D_{+i}^\ast D_{+j}
        +
        D_{\times i}^\ast D_{\times j}
    \right].
    \label{eq:appendix_metric_definition}
\end{equation}
The matrix is positive semidefinite and has rank four for the
benchmark configuration considered here.

For numerical conditioning and interpretation, each coordinate is
rescaled by its one-dimensional local
\(\rho_{T,{\rm leak,avg}}=5\) scale \(s_i\),
\begin{equation}
    u_i
    =
    \frac{\Delta\theta_i}{s_i}.
\end{equation}
The normalized metric \(\mathbf C\) is then defined by
\begin{equation}
    \rho_{T,{\rm leak,avg}}^2
    =
    25\,
    \mathbf u^T
    \mathbf C
    \mathbf u,
    \label{eq:appendix_normalized_metric}
\end{equation}
with \(C_{ii}=1\).
Its four resolved eigenvectors define the principal directions used
below.  The fifth mode is locally unresolved and is therefore not
assigned a finite principal-direction leakage scale.

The conservative worst-polarization leakage is evaluated separately.
For a given parameter displacement, the two tensor polarizations define
a \(2\times2\) Hermitian leakage-information matrix,
\begin{equation}
    \left(\mathbf H_{\rm leak}\right)_{pq}
    =
    \int dt\,
    \frac{
        \mathcal A^2(t)\,
        \ell_p^\ast
        \left(
            t;\Delta\boldsymbol{\theta}
        \right)
        \ell_q
        \left(
            t;\Delta\boldsymbol{\theta}
        \right)
    }{
        S_{\Delta\boldsymbol{\theta}}(t)
    },
    \qquad
    p,q=+,\times .
    \label{eq:appendix_leakage_matrix}
\end{equation}
The tensor normalization is described by the corresponding
\(2\times2\) information matrix of the unequal-arm \(A/E\) network,
\begin{equation}
    \left(\mathbf Q_T\right)_{pq}
    =
    \int dt\,
    \mathcal A^2(t)\,
    \left(
        R_{p,M}^{AE}
    \right)^*
    \left(
        C_{AE}^{-1}
    \right)^{MN}
    R_{q,N}^{AE},
    \qquad
    p,q=+,\times,
    \quad
    M,N\in\{A,E\}.
    \label{eq:appendix_tensor_information_matrix}
\end{equation}
The polarization-averaged tensor information used above satisfies
\begin{equation}
    I_T
    =
    \frac{1}{2}
    {\rm Tr}\,\mathbf Q_T.
    \label{eq:appendix_tensor_information_trace}
\end{equation}

Let \(\lambda_{\max}\) denote the largest generalized eigenvalue of
\begin{equation}
    \mathbf H_{\rm leak}\mathbf v
    =
    \lambda\,
    \mathbf Q_T\mathbf v.
    \label{eq:appendix_generalized_eigenvalue}
\end{equation}
The conservative worst-polarization leakage SNR is then
\begin{equation}
    \rho_{T,{\rm leak,worst}}
    =
    \rho_T
    \sqrt{\lambda_{\max}}.
    \label{eq:appendix_worst_leakage_snr}
\end{equation}
Thus the polarization-averaged leakage is represented by the
well-defined \(5\times5\) metric
\(\mathbf G_{\rm leak}\), whereas the worst-polarization result is
evaluated from the full tensor-polarization matrices rather than from
a separate \(5\times5\) metric.

For the local derivative model,
\(\ell_p\) is replaced by its linear expansion in
Eq.~\eqref{eq:appendix_leakage_derivative}, while
\(S_{\Delta\boldsymbol{\theta}}(t)\) is replaced by \(S_0(t)\).
Since the leakage amplitude is already first order in
\(\Delta\boldsymbol{\theta}\), the mismatch dependence of the noise
denominator contributes only beyond quadratic order.

To compare the local model with direct recomputation along a resolved
principal direction, let
\(\mathbf v_a\) be a unit eigenvector of \(\mathbf C\) with eigenvalue
\(\lambda_a>0\).  We parameterize the displacement along that direction
by
\begin{equation}
    \mathbf u(q)
    =
    \frac{q}{5\sqrt{\lambda_a}}\,
    \mathbf v_a,
    \label{eq:appendix_principal_coordinate}
\end{equation}
so that the local polarization-averaged prediction is simply
\begin{equation}
    \rho_{T,{\rm leak,avg}}^{\rm local}
    =
    |q|.
    \label{eq:appendix_principal_coordinate_snr}
\end{equation}
Thus \(|q|=5\) corresponds to the reference leakage level used in the
main text.

\begin{figure*}[t]
    \centering
    \includegraphics[
        width=0.92\textwidth
    ]{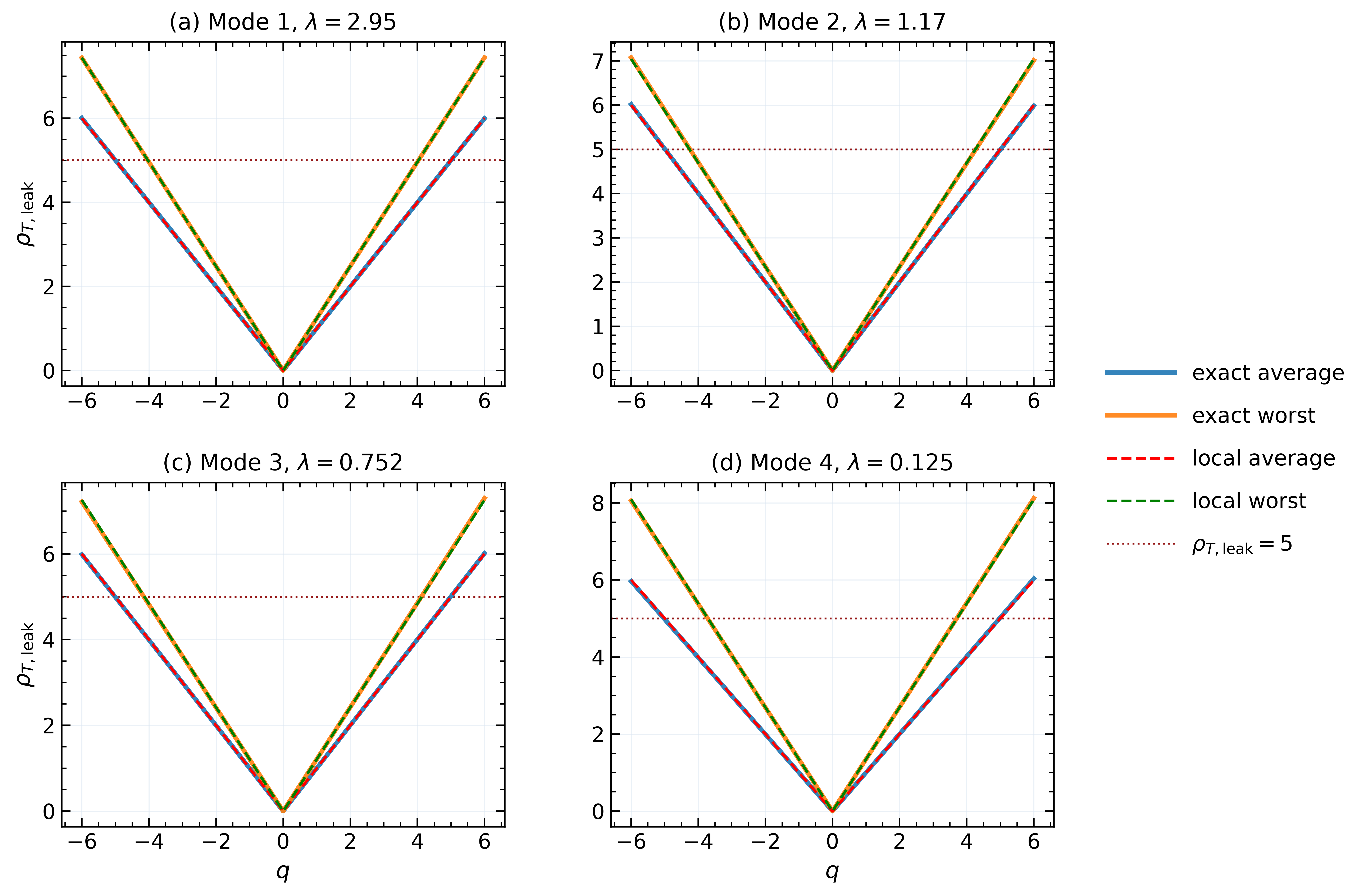}
    \caption{
    Validation of the local tensor-leakage model along the four
    resolved principal directions of the normalized joint metric.
    Solid curves show direct recomputation of the mismatched
    tensor-null response, while dashed curves show the corresponding
    local predictions.  Results are shown for both the
    polarization-averaged and worst-polarization tensor leakage.
    The principal-direction coordinate \(q\) is defined by
    Eq.~\eqref{eq:appendix_principal_coordinate}, so that the local
    polarization-averaged prediction satisfies
    \(\rho_{T,{\rm leak,avg}}^{\rm local}=|q|\).
    The fifth eigenvector is locally unresolved and therefore has no
    finite quadratic leakage scale.
    }
    \label{fig:appendix_principal_leakage_scans}
\end{figure*}

Figure~\ref{fig:appendix_principal_leakage_scans} compares the local
model with direct recomputation of the mismatched tensor-null response
along the four resolved principal directions.  Both the
polarization-averaged and worst-polarization leakages are shown.
The exact curves follow the local predictions closely in the
small-mismatch region relevant to the leakage thresholds used in the
main analysis.

The principal-direction comparison is supplemented by random joint
perturbations that are not aligned with the eigendirections of
\(\mathbf G_{\rm leak}\).  At small radius, the relative discrepancy
between the exact polarization-averaged leakage and the quadratic
metric prediction is
\begin{equation}
    4.336\times10^{-4},
    \label{eq:appendix_random_metric_error}
\end{equation}
while the corresponding worst-polarization diagnostic differs by
\begin{equation}
    9.797\times10^{-4}.
    \label{eq:appendix_random_worst_error}
\end{equation}
The agreement therefore remains at the sub-\(10^{-3}\) level away
from the principal axes.

Several independent numerical checks are summarized in
Table~\ref{tab:appendix_metric_validation}.  At zero mismatch the
residual tensor leakage is at the \(10^{-11}\) SNR level, well below
the leakage scales considered in the robustness analysis.  The
Richardson correction to the joint metric is below
\(5\times10^{-9}\).  Repeating the calculation on two independent
time samplings changes the normalized metric by less than
\(2\times10^{-5}\) in relative Frobenius norm and changes the resolved
eigenvalues by approximately \(1.4\times10^{-4}\).

\begin{table}[t]
\caption{
Numerical checks of the local joint tensor-leakage calculation.
The reported metric quantities refer to the polarization-averaged
\(5\times5\) metric.
}
\label{tab:appendix_metric_validation}
\centering
\begin{ruledtabular}
\begin{tabular}{l c}
check & value \\
\hline
zero-mismatch average leakage
&
\(2.088\times10^{-11}\)
\\
zero-mismatch worst leakage
&
\(2.398\times10^{-11}\)
\\
zero-mismatch scalar-retention error
&
\(1.066\times10^{-15}\)
\\
Richardson correction
&
\(4.662\times10^{-9}\)
\\
random-joint average-metric error
&
\(4.336\times10^{-4}\)
\\
stride-1/stride-2 metric difference
&
\(1.872\times10^{-5}\)
\\
stride-1/stride-2 resolved-eigenvalue difference
&
\(1.407\times10^{-4}\)
\end{tabular}
\end{ruledtabular}
\end{table}

These tests establish the local metric as an accurate description of
tensor leakage in the small-mismatch region used in the main text.
They do not turn
\(\mathbf G_{\rm leak}\) into a parameter-estimation Fisher matrix.
The metric quantifies how a specified error in the source parameters
propagates into residual tensor power when constructing the
source-tracked null channel; it does not provide a posterior covariance
or determine how accurately those parameters can themselves be
measured.

\clearpage
\bibliography{Ref}

\end{document}